\documentclass[11pt]{article}

\usepackage[T1]{fontenc}
\usepackage[utf8]{inputenc}
\usepackage[numbers]{natbib}
\usepackage[margin=1in]{geometry}
\usepackage{longtable}
\usepackage{hyperref}
\usepackage{amsmath}
\usepackage{authblk}
\usepackage{tikz}
\usetikzlibrary{arrows,shapes,positioning}
\usepackage[ruled,norelsize]{algorithm2e}

\title{\textbf{Spectrum-based Software Fault Localization: }\\ \textbf{A Survey of Techniques, Advances, and Challenges}}
\author[1]{Higor A. de Souza\thanks{hamario@ime.usp.br}}
\author[2]{Marcos L. Chaim\thanks{chaim@usp.br}}
\author[1]{Fabio Kon\thanks{fabio.kon@ime.usp.br}}
\affil[1]{Department of Computer Science -- University of S\~ao Paulo}
\affil[2]{School of Arts, Sciences, and Humanities -- University of S\~ao Paulo}
\date{}

\begin{document}
\maketitle

\begin{abstract}
Despite being one of the most basic tasks in software development, debugging is still performed  
in a mostly manual way, leading to high cost and low performance. To address this problem, 
researchers have studied promising approaches, such as Spectrum-based Fault Localization (SFL) 
techniques, which pinpoint program elements more likely to contain faults. This survey discusses 
the state-of-the-art of SFL, including the different techniques that have been proposed, the type and 
number of faults they address, the types of spectra they use, the programs they utilize in their validation, 
the testing data that support them, and their use at industrial settings. Notwithstanding the advances, 
there are still challenges for the industry to adopt these techniques, which 
we analyze in this paper. SFL techniques should propose new ways to generate reduced sets of 
suspicious entities, combine different spectra to fine-tune the fault localization ability, 
use strategies to collect fine-grained coverage levels from suspicious coarser levels for 
balancing execution costs and output precision, and propose new techniques to cope with 
multiple-fault programs. Moreover, additional user studies are needed to understand better how 
SFL techniques can be used in practice. 
We conclude by presenting a concept map about topics and challenges for future research in SFL.
\\
\end{abstract}

\section{Introduction}
Program faults are an inevitable consequence of writing 
code. Faults occur for various reasons: typing errors, 
misunderstanding software requirements, wrong values assigned to variables, 
or absence of code to cope with some unpredicted condition.
During the testing phase or in the field, a fault is revealed 
when a program presents an unexpected behavior (known as failure). 
Once a failure occurs, the two-step debugging process begins. First, a 
developer inspects the code to \textit{locate} the failure's cause. 
Second, s/he \textit{fixes} the fault \cite{myers1979}.

Fault localization is a costly activity in the software development process.
Testing and debugging can account for up to 75\% of development costs \cite{tassey2002}.
In practice, fault localization is performed manually; developers observe 
failing test cases and then search the source code for faults. 
They use their knowledge of the code to investigate excerpts 
that may be faulty. The most frequent debugging practices include inserting print 
statements and breakpoints, checking the stack trace, and verifying 
failing test cases. Since these manual processes can be expensive and ad-hoc \cite{jones2007}, 
approaches that automate fault localization are valuable for software development 
cost reduction.

Several techniques for automating fault localization have been proposed in the 
last decades \cite{agrawal1995,jones2002,wotawa2002,zeller2002,renieris2003}. 
These techniques use testing data to suggest which program entities, 
such as statements, predicates, definition-use associations, and call functions \cite{renieris2003}, 
are more likely to be faulty. 
Using fault localization results, developers can inspect the code 
to search for bugs guided by a set of suspicious entities.

\subsection{Motivation and scope}

This survey's main scope is to analyze fault localization techniques 
that use dynamic information from test execution, known as Spectrum-based 
Fault Localization (SFL) or Coverage-based Fault Localization. These 
techniques have achieved significant results when compared to other 
fault localization techniques and often present low overhead costs.

Despite the growing number of available SFL techniques, they 
are still unknown to practitioners. Many factors explain this.
In general, SFL techniques have been  evaluated using a set of known programs. In most cases, 
they are small and contain a single fault by version. 
In practice, though, developers tackle large programs with an unknown number of faults.
To complicate things, these bugs may interact and produce different failures depending
on the failing test case. User studies in which developers debug real faulty programs with the support of SFL techniques 
could shed light on such techniques' effectiveness and efficiency, but they are scarce. As a consequence, the existing 
SFL techniques are rarely used to automate software companies' debugging processes \cite{parnin2011}. 

This survey presents a comprehensive view of state-of-the-art SFL techniques proposed from 2005 to October 2017. 
It describes the most recent advances and challenges, which includes: approaches and testing data used 
by SFL techniques; the number and characteristics of faults; benchmark programs 
used in experiments; costs of SFL techniques; new ways to provide fault localization results to
developers; and practical use of SFL techniques.

We discuss and summarize features, limitations, and challenges, 
indicating future directions for improving SFL. 
The techniques are classified according to their debugging support 
strategies and relevance to the main topics.
We also present a concept map \cite{novak2008} of the SFL area, addressing the relationships among topics, 
their roles, and challenges for future research.
We believe the information in this survey is useful to both researchers interested in understanding and 
improving fault localization techniques, especially Spectrum-based Fault Localization, and practitioners interested 
in improving their debugging processes.

The remainder of this paper is organized as follows. Section~\ref{sec-overview} 
presents an overview of the fault localization area, including its history, 
and the terminology used in the area. We describe the scope, criteria to select relevant papers
for this survey and the main topics regarding fault localization in Section~\ref{sec-scope}. 
In Section~\ref{sec-techniques}, we present different fault localization approaches, 
focusing on spectrum-based techniques. Section~\ref{sec-spectra} shows the different spectra 
used by SFL techniques.
Concerns regarding faults are shown in Section~\ref{sec-faults}. The 
programs used to evaluate the techniques are presented in Section~\ref{sec-programs}. 
Section~\ref{sec-testing} shows the concerns related to the use of testing 
information in fault localization. The practical use of fault localization 
is presented in Section~\ref{sec-practical_use}. 
We discuss main challenges and future directions in Section~\ref{sec-discussion}. 
Related works are shown in Section~\ref{sec-related_works}.
Finally, we draw our conclusions in Section~\ref{sec-conclusion}.

\section{Concepts and seminal studies}
\label{sec-overview}

In this section, we define the main terms used by fault localization studies and present a historical overview of seminal works. 

\subsection{Terminology}

Due to the diversity of studies on fault localization, several terms have been used to define 
similar concepts. In what follows, we clarify terms and synonyms used in 
the studies addressed in the survey.

\subsubsection*{Faults, errors, and failures}

Faults, errors, and failures represent three stages in the execution of a program 
during which an unexpected behavior occurs. The IEEE Standard 610.12 
\cite{ieee1990} defines fault, error, and failure as follows. \textit{Fault} is an 
incorrect step, process, or data definition in a computer program. A fault is 
inserted by a developer who wrote the program. A fault is also called 
\textit{bug} or \textit{defect}.

\textit{Error} is a tricky term, which is also sometimes used to refer to a fault, 
failure, or mistake. In its particular sense, an error is a difference between 
a computed value and the correct value \cite{ieee1990}. The term is often used to 
indicate an incorrect state during program execution. Thus, an error 
occurs when an executed fault changes the program state. Other terms used to 
express error are \textit{infection} and \textit{anomaly}.

\textit{Failure} describes a system's inability to perform its function at the 
expected requirements \cite{ieee1990}. A failure is observed as an unexpected output, 
which occurs when an error in a program state leads to a wrong 
output. A synonym for failure is \textit{unexpected behavior}. \textit{Crashes} are 
failures that interrupt program executions and thus have an apparent behavior. 
\textit{Mistakes} are human actions that produce faults \cite{ieee1990}.

\subsubsection*{Real and seeded faults}

The literature on fault localization refers to two categories of faults. \textit{Seeded 
faults} are those intentionally inserted for monitoring detection \cite{ieee1990}. 
Faults can be manually inserted for experimental purposes, or by using mutation testing \cite{demillo1978}. 
Fault seeding is also known as \textit{fault injection}. Conversely, 
\textit{real faults} are those that naturally occur during software development. 

\subsubsection*{Ranking metrics}

\textit{Ranking metrics} are used in fault localization to calculate the likelihood that program 
entities will be faulty. The studies on fault localization use different terms to 
refer to ranking metrics: \textit{technique}, \textit{risk evaluation formula}, 
\textit{metric}, \textit{heuristic}, \textit{ranking heuristic}, 
\textit{coefficient}, and \textit{similarity coefficient}.

\subsubsection*{Program entity}

A \textit{program entity} is a part of a program. 
It comprises any granularity of a program, from a statement to a subprogram. 
Program entities include \textit{statements}, \textit{blocks}, \textit{branches}, 
\textit{predicates}, \textit{definition-use associations}, \textit{components}, 
\textit{functions}, \textit{program elements}, and \textit{program units}.

\subsubsection*{Spectrum-based fault localization}
Program spectra \cite{reps1997}, also known as code coverage, can be defined 
as a set of program entities covered during test execution \cite{renieris2003}. 
\textit{Spectrum-based fault localization} uses information from program 
entities executed by test cases to indicate entities more likely to be faulty. 
There are several synonyms of program spectrum used in the literature, such as 
\textit{code coverage}, \textit{testing data}, 
\textit{dynamic information}, \textit{execution trace}, \textit{execution path}, 
\textit{path profile}, and \textit{execution profile}.

\subsubsection*{Suspicious program entities}

Program entities more likely to contain faults are called \textit{suspicious}, 
\textit{suspected}, \textit{candidate}, and \textit{faulty elements}.

\subsection{A Brief History of Debugging Techniques}

Unfortunately, developing programs without making mistakes is nearly impossible. 
Therefore, debugging is an inherent programming activity.
The use of the word \textit{bug} originates in Thomas Edison's time. It was used to 
indicate flaws in engineering systems \cite{kidwell1998}. In the late 1940s, 
the \textit{Mark II} computer at \textit{Harvard University} suddenly stopped. 
Technicians found that a dead moth had shorted out some of the computer's circuits, 
and taped the bug into the machine's logbook \cite{kidwell1998}. 
The term \textit{debug} was then associated with the activities of finding and fixing 
program faults.
According to \citet{araki1991}, the most primitive debugging practice 
entails inserting print statements in the code to verify the state of variables. 

Despite advances, in practice fault localization has changed little over time. 
Most of the techniques used by developers today were proposed in the 
1960s \cite{agrawal1989}, and earlier debugging tools originate from the late 1950s \cite{evans1966}. 
Some examples are \textit{Gilmore's debugging tool} (called \textit{TX-O Direct Input Utility System}) \cite{gilmore1957}, 
\textit{FLex Interrogation Tape} (\textit{FLIT}) \cite{stockham1960}, 
and \textit{DEC (Digital Equipment Corporation) Debugging Tape} (\textit{DDT}) \cite{kokot1961}.
The TX-O Direct Input Utility System influenced subsequent, more 
sophisticated debugging tools. It was based on the idea of moving the debugging program to the computer's memory, 
making it possible to verify and modify registers during the execution. 
As the first tool to implement the concept of a breakpoint, FLIT also allows modifying 
the program during its execution. DDT evolved from FLIT for the \textit{PDP-1} computer. 
Another advance in the 1960s debugging tools was the conditional breakpoint, which 
permits a breakpoint's execution only when it reaches some specific condition.
The first tools to provide code tracing were those for high-level languages, such 
as debugging tools for Lisp and Prolog. Beyond code tracing, debugging tools for 
high-level languages do not present any additional features compared to those 
for assembly \cite{evans1966}. Indeed, the debugging tools used in 
industrial settings today do not differ much from the above-described techniques.

Nevertheless, several techniques have been proposed for automating debugging, most of them 
for fault localization. \citeauthor{balzer1969} presented a tool called 
\textit{EXtendable Debugging and Monitoring System} (\textit{EXDAMS}) \citep{balzer1969}, which was one of 
the first tools to allow either backward or forward navigation through the code. Its 
visualization provides control-flow and data-flow information using graphics, such as 
a tree structure of the execution at some point of interest. The execution data was stored 
in a history tape. However, EXDAMS did not use this information to suggest statements 
more likely to contain bugs, which would help developers in fault localization. 
\citeauthor{nagy1974} proposed one of the earliest techniques to automatically identify bugs 
by comparing successive versions of a program, considering that the 
changes in code are bug corrections \citep{nagy1974}. Thus, their technique provides bug patterns to aid debugging.

Some previous techniques tried to understand a program's whole behavior. These 
techniques depend on correct specifications of programs, which in practice are very 
difficult to obtain. \citeauthor{adam1980} proposed a tool called \textit{LAURA} \citep{adam1980}. The 
tool receives a program model represented by graphs. To identify faults, LAURA makes program 
transformations to compare  the original program with the program model. 
\citeauthor{johnson1985} proposed a tool called \textit{PROgram UnderSTanding} (\textit{PROUST}) \citep{johnson1985}, 
which receives \textit{programming plans}, the intentions that a developer has to write the code, 
and a program. PROUST has a knowledge base of programming plans that it compares to the input plan's goals. 
PROUST then generates a diagnostic output of bugs found in the code, including an explanation of 
mistakes causing the bugs. 

\textit{Assertions} are another strategy used to automate debugging. \citeauthor{fairley1979} 
proposed a debugging tool called \textit{Assembly Language Assertion Drive Debugging INterpreter} 
(\textit{ALADDIN}) \citep{fairley1979}, which uses breakpoint assertions instead of breakpoint locations. 
The breakpoint assertion is executed when a wrong value occurs in the program state at a certain 
point. Assertions can be helpful for detecting errors in some circumstances, especially for functions that 
calculate values or must contain a certain number of elements. However, it is not always possible 
to use assertions to detect all incorrect program behaviors, which may be infeasible for large and 
complex programs.

\citeauthor{shapiro1983} proposed two algorithms to detect incorrect program procedures \citep{shapiro1983}: 
one for deterministic programs, and another for non-deterministic programs. These algorithms 
ask an oracle if the output for a given input to a procedure is correct, repeating the process 
for the following program execution procedures. The first procedure with an incorrect 
output is the incorrect procedure.
The algorithms proposed by \citet{shapiro1983} suppose that developers perform the role of the oracle 
for each executed procedure, which may be an error-prone and time-consuming activity for long-running 
and large programs. 
The author suggests that a possible way to automate the oracle is to accumulate a knowledge database 
of developers' answers. \citeauthor{fritzson1992} implements such an idea using category partition testing \citep{fritzson1992}. 

\citeauthor{korel1988b} proposed a tool called \textit{Program Error-Locating Assistant System} 
(\textit{PELAS})---the first fault localization technique based on program dependence \citep{korel1988b}. PELAS 
asks developers about a behavior's correctness and uses the answer to 
indicate possible fault locations. Program dependence data is used to guide the 
developer navigation in searching possible fault sites. The author states that the backtracking 
reasoning used (based on program dependence) is an abstraction of experienced developers' 
intuitive processes. PELAS can narrow the amount of code a developer must verify. 

\textit{Program slicing}, a technique based on program static information, 
was proposed by \citet{weiser1981}. The technique generates subsets of a program, called slices, 
which contain the expected program behavior. Thus, a developer focuses his/her
attention on a reduced part of a program. Program slicing can be used for debugging, or to 
change the code, depending on the specification of 
program elements or variables of interest, called \textit{slicing criterion}. 
A slice is composed of elements that relate to such a criterion.
\citeauthor{korel1988a} proposed \textit{dynamic slicing} to reduce slice size \citep{korel1988a}. 
Dynamic slices are composed of only executed statements.
Although dynamic slicing reduces the amount of code to be inspected, the remaining code 
is still excessive, which makes it impractical. 

The use of testing data for fault localization has grown over the 
last few decades. 
\citeauthor{collofello1987} proposed the first fault localization technique that 
uses paths executed by tests to indicate faulty sites \citep{collofello1987}. They used ten ranking metrics to 
calculate the likelihood that program elements will be faulty, wherein a program element is a 
decision-to-decision path (DD-path)---the code chunk between two predicates.
The technique uses a set of DD-paths from passing test cases, and DD-paths from a single 
failing test case, to indicate DD-paths likely to contain faults. 
\citeauthor{agrawal1995} proposed \textit{execution dices} for fault localization \citep{agrawal1995}. 
An execution dice is a set of program basic blocks formed by the difference between two 
dynamic slices---one from a failing test case and the other one from a passing test case. 
Even reducing the amount of code returned, the dices still contain a large number of blocks 
for inspection. 

Other approaches were proposed in the early 2000s. 
\citeauthor{jones2002} present a technique called \textit{Tarantula} that uses a homonym ranking metric 
to calculate statements' suspiciousness \citep{jones2002}. 
The suspiciousness values are calculated according to the frequency of the statements in passing and 
failing test cases. These statements are classified and shown in a graphic visualization form, 
using different colors according to their suspiciousness values.
\citeauthor{zeller2002} applied the \textit{Delta Debugging} (DD) algorithm to find causes of failures that 
occur during execution, using the difference in program states (variables and values) between one 
passing and one failing run \citep{zeller2002}. 
These differences are reduced to obtain a minimal set that causes the 
failure. Such a subset is deemed the failure's cause. 
DD differs from other works by using program states instead of program elements.

Some techniques proposed for fault localization use information from static analysis.
These techniques are independent of testing and can be used to inspect all paths in a program. Static analysis 
advantageously assures that a program is fault-free by exploring all its possible interactions.
However, the performance of these techniques is tied to formal proofs of program correctness. 
Such proofs are infeasible in practice, even for small general purpose programs.
\citeauthor{wotawa2002} used \textit{Model-Based Diagnosis} (MBD), which was originally used for electronic digital circuits, 
for debugging software faults \citep{wotawa2002}. MBD generates a logical model from the source code and uses logical reasoning 
to obtain a minimal set of statements that explain\footnote{\textit{Explain} means components that are logically related to faulty behaviors.} the existing faults. 
\citeauthor{hovemeyer2004} proposed a tool that uses bug patterns from Java to locate bugs automatically \citep{hovemeyer2004}. 
The tool statically analyzes a program to search for bug patterns; they proposed fifty bug patterns. 
The technique generates a list of warnings (statements that might be faulty). \citeauthor{rutar2004} 
discuss tools that use static analysis to automatically locate bugs \citep{rutar2004}.

In this brief historical overview, we described many different approaches to improve the localization of program faults. 
However, print statements and symbolic debuggers are prevalent in practice. What is the reason for such a state of affairs in debugging?
Primarily, many of the techniques do not scale to programs developed in industry. Another reason is that 
the techniques are not assessed \textit{in situ}; that is, in real debugging situations. \citeauthor{parnin2011} showed
that the developer behavior might differ from that expected by the proponents of localization technique \citep{parnin2011}. 

The rest of this survey is dedicated to Spectrum-based Fault localization. SFL utilizes testing data to highlight suspicious
pieces of code. By using data already collected during testing, SFL tends to have lower overhead in comparison
to other debugging techniques. We discuss the more relevant results and challenges to industry adoption of SFL in the following sections.

\section{Selection of studies and scope}
\label{sec-scope}
This survey presents a comprehensive overview focused on techniques that use SFL. 
We searched for studies published from 2005 to October 2017 in the following digital libraries: \textit{ACM Digital Library}, 
\textit{IEEE Xplore Digital Library}, \textit{SpringerLink}, and \textit{SciVerse Scopus - Elsevier}. 

The studies included in this survey were published in journals and conferences with acknowledged 
quality. We combined database searches with \textit{backward snowballing} \cite{jalali2012} 
to expand the search for relevant results. 
We selected studies that proposed new techniques to perform fault 
localization based on program spectrum data. Only works that carried out an experimental evaluation of the proposed technique 
were included. We also considered studies that compare existing fault 
localization techniques, that propose improvements to program spectra data (e.g., testing information), and that assess practical use of SFL techniques. 
When available, only the extended versions of the papers were analyzed.

We selected papers after reading the title and abstract of all studies returned. 
In doubtful cases, other sections of the papers were read to decide whether they should be 
included. Selected papers were read in full. 
For the backward snowballing process, we first selected papers based on the description of such works from source 
papers. Then, we applied the same criteria to select the most relevant studies. 

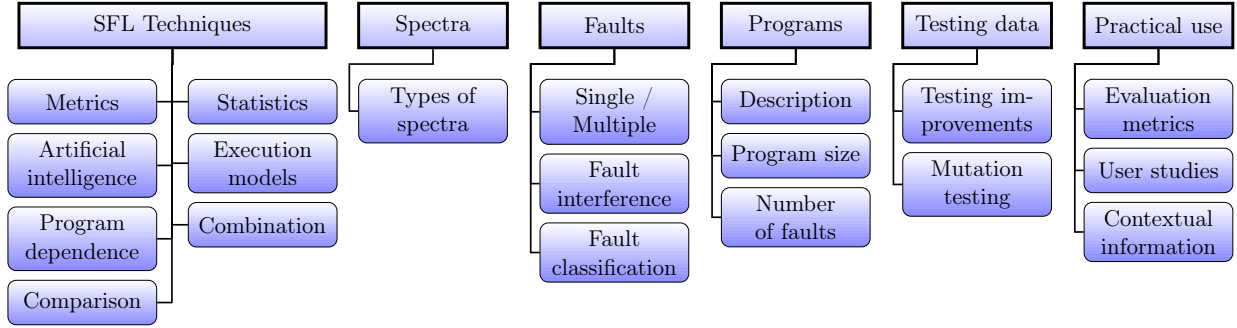
\begin{figure}
\resizebox {\columnwidth}{!}{
\begin{tikzpicture}[node distance=0.15cm,auto]
\tikzset{
  flnode/.style={rectangle,rounded corners,draw=black,text width={width("of techniques")+2pt}, text centered,minimum height=0.75cm,top color=white, bottom color=blue!50},
  flnodehead/.style={rectangle,thick,draw=black,text width={width("of techniques")+2pt}, text centered,minimum height=0.75cm,top color=white, bottom color=blue!30,line width=0.5mm},
  flnodeX/.style={rectangle,thick,draw=black,text width={140pt}, text centered,minimum height=0.75cm,top color=white, bottom color=blue!30,line width=0.5mm},
  flarrow/.style={-,>=latex',shorten >=1pt,thick},
  flcaption/.style={text width=7em, text centered},
  flline/.style={-,thick}
}
\node[flnodehead](bench){Programs};
\node[flnodehead,left=0.5cm of bench](fault){Faults};
\node[flnodehead,left=0.5cm of fault](spec){Spectra};
\node[flnodeX,left=0.5cm of spec](tech){SFL Techniques};
\node[below=0.75cm of tech](hiddentech){};
\node[flnodehead,right=0.5cm of bench](test){Testing data};
\node[flnodehead,right=0.5cm of test](pract){Practical use};
\node[flnode,below=0.5cm of bench](desc){Description};
\node[flnode,below=of desc](size){Program size};
\node[flnode,below=of size](nofaults){Number of faults};
\node[flnode,below=0.5cm of fault](singmult){Single / Multiple};
\node[flnode,below=of singmult](interf){Fault interference};
\node[flnode,below=of interf](faulttype){Fault classification};
\node[flnode,below=0.5cm of spec](typspec){Types of spectra};
\node[flnode,right=0.125cm of hiddentech](stat){Statistics};
\node[flnode,below=of stat](exec){Execution models};
\node[flnode,below=of exec](comb){Combination};
\node[flnode,left=0.125cm of hiddentech](met){Metrics};
\node[flnode,below=of met](artif){Artificial intelligence};
\node[flnode,below=of artif](dep){Program dependence};
\node[flnode,below=of dep](comp){Comparison};
\node[flnode,below=0.5cm of test](improv){Testing improvements};
\node[flnode,below=of improv](mut){Mutation testing};
\node[flnode,below=0.5cm of pract](eval){Evaluation metrics};
\node[flnode,below=of eval](exp){User studies};
\node[flnode,below=of exp](cont){Contextual information};

\draw[flarrow](bench.south) -- ++(0,0) -- ++(0,-0.25) -| ([yshift=-0.047cm,xshift=-0.15cm]desc.west);
\draw[flarrow](bench.south) -- ++(0,0) -- ++(0,-0.25) -| ([yshift=-0.047cm,xshift=-0.15cm]size.west);
\draw[flarrow](bench.south) -- ++(0,0) -- ++(0,-0.25) -| ([yshift=-0.047cm,xshift=-0.15cm]nofaults.west);
\draw[flarrow](tech.south) -- ++(0,0) -- ++(0,-0.25) -| ([yshift=-0.047cm,xshift=0.25cm]met.east);
\draw[flarrow](tech.south) -- ++(0,0) -- ++(0,-0.25) -| ([yshift=-0.047cm,xshift=0.25cm]artif.east);
\draw[flarrow](tech.south) -- ++(0,0) -- ++(0,-0.25) -| ([yshift=-0.047cm,xshift=0.25cm]dep.east);
\draw[flarrow](tech.south) -- ++(0,0) -- ++(0,-0.25) -| ([yshift=-0.047cm,xshift=0.25cm]comp.east);
\draw[flarrow](spec.south) -- ++(0,0) -- ++(0,-0.25) -| ([yshift=-0.047cm,xshift=-0.15cm]typspec.west);
\draw[flarrow](fault.south) -- ++(0,0) -- ++(0,-0.25) -| ([yshift=-0.047cm,xshift=-0.15cm]singmult.west);
\draw[flarrow](fault.south) -- ++(0,0) -- ++(0,-0.25) -| ([yshift=-0.047cm,xshift=-0.15cm]faulttype.west);
\draw[flarrow](fault.south) -- ++(0,0) -- ++(0,-0.25) -| ([yshift=-0.047cm,xshift=-0.15cm]interf.west);
\draw[flarrow](test.south) -- ++(0,0) -- ++(0,-0.25) -| ([yshift=-0.047cm,xshift=-0.15cm]improv.west);
\draw[flarrow](test.south) -- ++(0,0) -- ++(0,-0.25) -| ([yshift=-0.047cm,xshift=-0.15cm]mut.west);
\draw[flarrow](pract.south) -- ++(0,0) -- ++(0,-0.25) -| ([yshift=-0.047cm,xshift=-0.15cm]eval.west);
\draw[flarrow](pract.south) -- ++(0,0) -- ++(0,-0.25) -| ([yshift=-0.047cm,xshift=-0.15cm]exp.west);
\draw[flarrow](pract.south) -- ++(0,0) -- ++(0,-0.25) -| ([yshift=-0.047cm,xshift=-0.15cm]cont.west);

\draw[flline](desc.west) -- ++(-0.15,0);
\draw[flline](size.west) -- ++(-0.15,0);
\draw[flline](nofaults.west) -- ++(-0.15,0);
\draw[flline](singmult.west) -- ++(-0.15,0);
\draw[flline](faulttype.west) -- ++(-0.15,0);
\draw[flline](interf.west) -- ++(-0.15,0);
\draw[flline](typspec.west) -- ++(-0.15,0);
\draw[flline](stat.west) -- ++(-0.25,0);
\draw[flline](exec.west) -- ++(-0.25,0);
\draw[flline](comb.west) -- ++(-0.25,0);
\draw[flline](met.east) -- ++(0.25,0);
\draw[flline](artif.east) -- ++(0.25,0);
\draw[flline](dep.east) -- ++(0.25,0);
\draw[flline](comp.east) -- ++(0.25,0);
\draw[flline](improv.west) -- ++(-0.15,0);
\draw[flline](mut.west) -- ++(-0.15,0);
\draw[flline](eval.west) -- ++(-0.15,0);
\draw[flline](exp.west) -- ++(-0.15,0);
\draw[flline](cont.west) -- ++(-0.15,0);

\end{tikzpicture}
}
\caption{Fault localization topics}
\label{fig-fl_topics}
\end{figure}

We classified the papers regarding the research topics and challenges that characterize these studies. Figure~\ref{fig-fl_topics} shows 
the taxonomy that represents the main challenges addressed in this survey, which were 
classified into the following major topics: \textit{SFL Techniques}, \textit{Spectra}, \textit{Faults}, \textit{Programs}, \textit{Testing data}, 
and \textit{Practical use}. 
Most of the studies intersect more than one topic. For each topic, we present studies that present the 
most distinguishable contributions.
Sections~\ref{sec-techniques}~to~\ref{sec-practical_use} present the studies according to the topics presented above.

\section{Spectrum-based Fault Localization techniques}
\label{sec-techniques}

Spectrum-based Fault Localization techniques propose several strategies to 
pinpoint faulty program entities. Most of them rank suspicious entities by using 
ranking metrics or statistical techniques. Artificial intelligence approaches are also used for fault 
localization. Other SFL techniques are based on execution models that indicate suspicious 
entities by comparing passing and failing executions. Some studies combine previous 
techniques, while others compare the effectiveness of several techniques.

There are SFL techniques that make use of other program analysis information, 
like program dependencies, execution graphs, and clustering of program entities.
In this section, we present SFL techniques, exploring the strategies mentioned above 
to improve fault localization.

\subsection{Metric-based techniques}
\label{sub-heuristics}

Metric-based techniques are those that use ranking metric formulas to pinpoint faulty program entities. 
Most of the SFL techniques use or propose ranking metrics to improve fault localization.
To determine correlations between program entities and test case results, 
these ranking metrics use program spectrum information derived from testing as input. 
Each program entity receives a suspiciousness score that indicates how likely it is to be faulty. 
The rationale is that program entities frequently executed in failing 
test cases are more suspicious. Thus, the frequency in which entities are executed 
in failing and passing test cases is analyzed to calculate its suspiciousness score. 
There are ranking metrics specifically created for fault localization, and other 
metrics were adapted from areas such as molecular biology. 
Some studies perform experiments comparing ranking metrics. 
Works that propose or use ranking metrics are presented hereafter.

\textit{Tarantula} \cite{jones2002} was one of the first techniques proposed for SFL. 
Its formula is shown in Table~\ref{tb-heuristic}. 
For each program entity, Tarantula calculates the frequency in which a program entity is executed 
in all failing test cases, divided by the frequency in which this program entity is executed 
in all failing and passing test cases. 
Tarantula has been used in several studies. \citeauthor{jones2007} use Tarantula 
to calculate the suspiciousness score of statements for their parallel debugging 
technique \citep{jones2007} (see Subsection~\ref{sub-single_multiple}). 
\citeauthor{ali2009} use Tarantula to evaluate characteristics that can 
influence fault localization techniques \citep{ali2009}. 
Furthermore, Tarantula has been used as a benchmark by several 
other techniques \cite{wong2010,xie2013a}.
Table~\ref{tb-heuristic} shows some of the main 
ranking metrics used for SFL. We use the following nomenclature in Table~\ref{tb-heuristic}: 
$c_{ef}$ indicates the number of times a program entity ($c$) is executed ($e$) in 
failing test cases ($f$), $c_{nf}$ is the number of times a program entity is not 
executed ($n$) in failing test cases, $c_{ep}$ is the number of times a 
program entity is executed by passing test cases ($p$) and $c_{np}$ represents the number 
of times a program entity is not executed by passing test cases.

\begin{center}
\begin{table}[b]%
\centering
\caption{Ranking metrics for fault localization\label{tb-heuristic}} 
\scalebox{0.925}{
\begin{tabular*}{500pt}{@{\extracolsep\fill}cc|cc@{\extracolsep\fill}}%
  \hline
  \textbf{Ranking metric} & \textbf{Formula} & \textbf{Ranking metric} & \textbf{Formula} \\
  \hline
  	 Tarantula & {\Large $\frac{\frac{c_{ef}}{c_{ef} + c_{nf}}}{\frac{c_{ef}}{c_{ef} + c_{nf}} + \frac{c_{ep}}{c_{ep} + c_{np}}}$} & Ochiai & {\Large $\frac{c_{ef}}{\sqrt{(c_{ef} + c_{nf})(c_{ef} + c_{ep})}}$}\\
  \hline Jaccard & {\Large $\frac{c_{ef}}{c_{ef} + c_{nf} + c_{ep}}$} & Zoltar & {\Large $\frac{c_{ef}}{c_{ef} + c_{nf} + c_{ep} + \text{\footnotesize{10000}} \cdot \frac{c_{nf} c_{ep}}{c_{ef}}}$}\\
  \hline $O^p$ & {\Large $c_{ef} - \frac{c_{ep}}{c_{ep} + c_{np} + 1}$} & $O$ & $-1 \text{ if } c_{nf} > 0 \text{, otherwise } c_{np}$ \\
  \hline Kulczynski2 & {\Large $\frac{1}{2}\left ( \frac{c_{ef}}{c_{ef} + c_{nf}} + \frac{c_{ef}}{c_{ef} + c_{ep}} \right )$} & McCon & {\Large $\frac{c_{ef}^2 - c_{nf}c_{ep}}{(c_{ef} + c_{nf})(c_{ef} + c_{ep})}$} \\
  \hline DStar & {\Large $\frac{c^{*}_{ef}}{c_{nf} + c_{ep}}$} & Minus & {\normalsize $\frac{\frac{c_{ef}}{c_{ef} + c_{nf}}}{\frac{c_{ef}}{c_{ef} + c_{nf}} + \frac{c_{ep}}{c_{ep} + c_{np}}} - \frac{1 - \frac{c_{ef}}{c_{ef} + c_{nf}}}{1 - \frac{c_{ef}}{c_{ef} + c_{nf}} + 1 - \frac{c_{ep}}{c_{ep} + c_{np}}}$} \\
  \hline
\end{tabular*}}
\end{table}
\end{center}

Some studies propose ranking metrics similar to Tarantula for techniques 
that use other coverage types. \citeauthor{masri2010} uses a \textit{Tarantula-like} 
ranking metric for the DIFA coverage \citep{masri2010} (see Section~\ref{sec-spectra}). This 
ranking metric is combined with another that considers only the percentage of executions 
in failing test cases. The final suspiciousness score averages the two. 
\citeauthor{wang2009} use a ranking metric similar to Tarantula for 
basic blocks \citep{wang2009}. \citeauthor{chung2008} propose a ranking metric for predicates that is 
also similar to Tarantula \citep{chung2008}. 

In addition to Tarantula, other techniques based on similarity 
formulas have been proposed in the last years. \citeauthor{abreu2009c} propose 
using \textit{Ochiai} and \textit{Jaccard} similarity coefficients as 
fault localization ranking metrics \citep{abreu2009c}. Ochiai was originally used in 
molecular biology, and Jaccard was used by \citeauthor{chen2002} to indicate 
faulty components in distributed applications \citep{chen2002}. The results of \citet{abreu2009c} show that both 
Ochiai and Jaccard outperform Tarantula on fault localization effectiveness. 
From these results, several works have used Ochiai \cite{santelices2009,digiuseppe2014}. 
Jaccard was not used on its own by any study presented in this 
survey. However, it was used in studies that compare the performance of several 
ranking metrics \cite{naish2010,xie2013a,ma2014}. 
Ochiai and Jaccard do not take into account program entities that are not executed in passing 
test cases ($c_{np}$). Thus, these ranking metrics assign more discriminative suspiciousness values 
for the well-ranked entities than those suspiciousness values assigned by Tarantula.
The ranking metric \textit{Zoltar} proposed by \citet{gonzalez2007} is a 
variation of \textit{Jaccard}. Its formula increases, even more, the suspiciousness of 
program entities that are more frequently executed in failing test cases. Zoltar was developed 
to detect errors in embedded systems using program invariants. \citet{abreu2009a} and 
\citet{abreu2011} use \textit{Zoltar} for fault localization.

\citeauthor{xu2013} propose a Tarantula-like ranking metric called \textit{Minus KBC} (MKBC) 
for their KBC coverage \citep{xu2013} (see Section \ref{sec-spectra}). 
The difference is that MKBC subtracts the percentage that a KBC is not 
executed in failing test cases, and divides by the percentage that such KBC 
is not executed for all executions (see Table~\ref{tb-heuristic}). 
This complementary frequency is called \textit{noise}; it decreases 
the importance of non-execution in the analysis. 

\citeauthor{naish2011} propose two new ranking metrics optimized for single-fault programs, 
called \textit{O} and \textit{$O^p$} \citep{naish2011}. Assuming the existence of a 
single fault, only statements that are executed in all failing test cases are fault candidates. 
\textit{Kulczynski2} (from Artificial Intelligence) and \textit{McCon} 
(from studies of plankton communities) are ranking metrics that have presented better 
results for programs with two simultaneous faults \cite{naish2009}.
\citeauthor{wong2012c} presented a technique called \textit{DStar} \citep{wong2012c}, which is a modified version of the Kulczynski 
ranking metric. The idea behind DStar is that the execution trace of each statement through test cases can be 
viewed as an execution pattern. Thus, the similarity between statements more frequently executed by failing 
test cases can pinpoint the faulty ones. 

Other techniques propose different strategies in conjunction with ranking metrics. 
\citeauthor{jeffrey2008} presented a technique that replaces values of statements in failing executions \citep{jeffrey2008}. If a 
failing execution then passes, such statement is classified between the most suspicious. The technique uses 
values observed in a statement from all executions, and only one value for each statement is replaced per execution.
\citeauthor{naish2009} presented an approach that assigns different weights 
to statements in failing test cases according to the number of statements in an 
execution \citep{naish2009}. The lower the number of commands, the greater the chance that one of 
them will be faulty. 

\citeauthor{xie2010a} proposed a technique that forms two groups of statements: one with 
statements that the failing test cases executed at least once, and another 
with statements that these test cases did not execute \citep{xie2010a}. The first group's statements 
receive suspiciousness scores.
\citeauthor{bandyopadhyay2011a} proposed a technique that assigns different weights 
for test cases according to the similarity of a passing and a failing 
test case \citep{bandyopadhyay2011a}. The more similar a passing test case is, the higher is its weight. The rationale 
is that a fault is more distinguishable since that there are few differences between the 
failing and the passing test cases. 

\subsection{Statistics-based techniques}
\label{sub-statistics}

Statistical techniques have also been applied in fault localization, 
and are used in the same manner as metric-based techniques. However, each 
statistical technique uniquely deals with testing information.
\citeauthor{liblit2005} proposed using conditional probability to evaluate 
predicates during program executions for fault localization \citep{liblit2005}. Predicates that 
are evaluated only as true in failing executions are deemed as more suspicious.
They calculate the probability that a predicate \textit{p} with value true causes 
a failure (\textit{Failure(p)}), and the probability that an execution of \textit{p} 
causes the failure (\textit{Context(p)}). The difference between such values 
(\textit{Importance}) indicates the suspiciousness of each predicate.
\citeauthor{liu2006a} proposed a statistical fault localization technique called SOBER \citep{liu2006a}
that considers the results of predicates evaluated as true multiple times per execution 
for several executions. They used \textit{Bernoulli distribution} to model each 
predicate's result for each execution. Then, conditional probability for passing and failing 
executions is used to calculate a predicate's bug relevance. 

\citeauthor{nainar2007} proposed the idea of complex predicates as bug predictors for fault localization \citep{nainar2007}. Complex predicates are formed 
by two predicates that are evaluated in each execution using boolean function operations (conjunction and disjunction). They argued 
that combinations of predicates can enhance fault localization when the predicates that form the complex predicates are already good bug predictors.
\citeauthor{baah2010} studied the use of causal inference for fault localization \citep{baah2010}, aiming to enhance fault localization by isolating causal effects that 
occur between program entities in a program dependence graph. This isolation can improve 
the values assigned to faulty entities by reducing noise caused by other program elements in the presence of failures. 
They used causal graphs concepts to identify entities that are independent in a program dependence graph. They also 
proposed a linear regression model to calculate the causal effect of statements on failures.

\citeauthor{zhang2011a} proposed a technique for using a non-parametric statistical 
model for fault localization \citep{zhang2011a}. They observed that the distribution of predicates 
during executions is non-normal (see Section~\ref{sec-spectra}). 
To calculate the suspiciousness of predicates, they consider the 
difference between the probability density function of a random variable of 
passing test cases and a failing test case.
\citeauthor{wong2012a} use a crosstab-based technique for fault localization \citep{wong2012a}.
The chi-square test was used to calculate the null hypothesis that an execution 
is independent of the coverage of a statement. Chi-square is applied to deal with 
categorical variables; in this work, such values are the test case results (pass 
and fail), and the presence of a statement in an execution (executed or not). Thus, they 
measured the dependency between an execution and a statement.

\citeauthor{zhang2012} used maximum likelihood estimation and linear regression to tune 
in SFL lists \citep{zhang2012}. They used previously known lists and bug positions, assuming a 
symmetric distribution of bug positions in such lists. Thus, they estimated a 
position to adjust the lists.
\citeauthor{xue2013a} applied the \textit{odds ratio} to SFL \citep{xue2013a}. The odds ratio is a 
statistical technique often used in classification and text mining problems. 
The odds ratio measures the strength or the weakness of a variable 
(a statement executed or not executed) associated with an event 
(a passing or a failing test case).
\citeauthor{xu2016} proposed the use of probabilistic cause-effect graphs (PCEG), which is 
obtained using program dependence graph of failing test cases \citep{xu2016}. The technique 
applies PCEG (a Bayesian network variation) to calculate the probability of statements to be faulty. 
The most suspicious statements are those more strongly connected to other suspicious 
statements.

\subsection{Artificial intelligence-based techniques}
\label{sub-ai}

Artificial Intelligence\footnote{We are considering Artificial Intelligence as a broader area, 
which includes Data Mining and Machine Learning techniques.} (AI) techniques have been applied for fault localization, 
using program spectrum data as input for classifying suspicious program elements.
\citeauthor{liu2005b} proposed a technique that uses \textit{graph mining} and \textit{support vector machines} (SVM) for fault 
localization \citep{liu2005b}. Program executions are represented as behavior graphs. Each node of a behavior graph is an executed function. 
The graphs are labeled with their execution result. 
Graph mining is applied to discover frequent subgraphs, which are features used to classify the graphs. 
SVM is applied to classify incorrect and correct executions. 
Thus, a sequence of bug-relevant functions is presented for a developer's inspection.
\citeauthor{nessa2008} applied \textit{N-gram analysis}, from Data Mining, for fault localization \citep{nessa2008}. The technique 
generates N-grams, subsequences of statements, from program spectra. The technique uses 
\textit{Association Rule Mining} to calculate the conditional probability that each N-gram relates to faulty 
executions. A list of the most suspicious statements is obtained from the most suspicious N-grams.

The technique proposed by \citet{murtaza2008} uses a \textit{decision tree} as a 
heuristic to indicate the origin of a fault. The technique identifies 
patterns of
function calls related to the fault.
\citeauthor{wong2012b} developed a \textit{Radial Basis Function neural network} (RBF-NN) 
for fault localization \citep{wong2012b}. The RBF-NN is modeled as statements representing the input 
neurons. 
Thus, the neural network acts as a ranking metric, in which the neural network 
individually evaluates each statement to determine its suspiciousness.
\citeauthor{lucia2014} assessed the use of association measures for fault localization \citep{lucia2014}. They evaluated 20 association measures 
commonly used in data mining and statistics, such as \textit{Yule-Q}, \textit{Yule-Y}, and \textit{odds ratio}. 
They modeled the problem as the strength of association between each entity's execution (and non-execution) 
and failures. Two association measures had a performance comparable to Ochiai: \textit{information gain} 
and \textit{cosine}. However, none of these association measures resulted in significant improvements for SFL. 

\citeauthor{roychowdhury2011b} proposed a technique based on \textit{Feature Selection} from Machine Learning \citep{roychowdhury2011b}.
They used two well-known methods, \textit{RELIEF} and \textit{RELIEF-F}, to classify statements according to their 
relevant potential to be faulty. The coverage matrix obtained from the test execution is applied to RELIEF 
and RELIEF-F. Each executed statement is considered a feature, and each coverage of a certain test case 
is a sample. Thus, the technique captures the diversity of these statements' behaviors as they relate to bug execution.
\citeauthor{zhang2014} applied \textit{Markov Logic Network} from machine learning to fault localization \citep{zhang2014}. They modeled the 
problem by considering dynamic information (program spectra), static information (control-flow and data-flow dependence), 
and prior bug knowledge (locations of similar bugs found in the past). The technique is developed for single-bug programs. 

\subsection{Execution models}
\label{sub-execution-model}

Another strategy used by SFL techniques is to build execution models from test executions.
Such models represent patterns of failing and/or passing executions.
Execution models are used to identify entities that meet or flee from 
an expected pattern. There are also models represented by graphs of execution.
\citeauthor{wang2005} proposed a technique that generates a passing run from a failing run \citep{wang2005}. The technique consists of toggling the results of branches 
in the failing run until obtaining a passing run. The outcome is a list of branches that were modified in the passing run. The toggling 
process starts on the last branch executed, and more branches are incrementally toggled until a passing run is located. A developer must 
check whether the generated run is correct.
\citeauthor{zhang2006} also proposed a technique that toggles branches of a failing run to generate a passing one \citep{zhang2006}. However, only one predicate is 
toggled per execution. The process starts on the last executed predicate and can be applied backward in all the following predicates until 
generating a passing run. 

The technique proposed by \citeauthor{zhang2009a}, called \textit{Capture Propagation} \citep{zhang2009a}, 
performs the propagation of suspicious values between related code blocks using 
graph models. The technique generates two mean edge profiles, one for all passing 
test cases, and another for all failing test cases. These profiles are used to 
obtain the suspiciousness for each edge. 
A propagation ratio of an edge is calculated according to the number 
of edges that enter in a successor block. Finally, the suspiciousness value 
of the predecessor block is obtained from the sum of the propagation ratio of 
all successor blocks for which the predecessor block has an edge.
The technique presented by \citet{mariani2011}, detailed in Section 
\ref{sec-spectra}, uses a behavior model of method calls from passing test cases. 
This model is compared with interactions from failing test cases to indicate suspicious 
interactions. \citeauthor{liu2010} proposed a technique in which the model 
is composed of messages from objects within the same class \citep{liu2010}. 

\citeauthor{wan2012} proposed a behavioral model that is constructed using two different coverage types:  
objects of an OO program and calls occurred inside each object \citep{wan2012}. A model is a sequence of objects and calls 
in execution traces of failing and passing test cases. The model contains two levels: the objects, and its internal calls. 
Two values from each entity are used to compose a suspicious value: coverage and violation. The violation means that entities 
from failing executions that are not present in the model are more suspicious, and entities which are present in the model from 
passing executions are less suspicious.
\citeauthor{dandan2014} proposed a probabilistic model of state dependency for fault localization \citep{dandan2014}. State dependency relates to 
predicate statement, which can be true or false. The technique calculates the probability that a predicate is true or false in 
passing and failing executions. Two probability models are generated: one for passing executions, and another for failing executions. 
The probability of control dependent statements is based on the probability of their parent statements. The suspiciousness value 
for each statement is then calculated using the models' probability values, and the outcome is a list of the most suspicious 
statements.

\citeauthor{laghari2016} proposed a technique that indicates methods more likely to be faulty \citep{laghari2016}. The technique 
uses information from integration testing, capturing patterns of method calls for each executed method through the use 
of \textit{closed itemset mining} algorithm. 
These methods are ranked according to the highest score of its patterns.

\subsection{Program dependence-based techniques}
\label{sub-program-dependence}

In their attempts to highlight faulty code excerpts, some 
techniques use additional program analysis information 
to highlight faulty code excerpts, such as 
program dependence data, program slicing, and 
assignment of weights to differentiate program entities.
The technique proposed by \citeauthor{zhang2009a} (see Subsection~\ref{sub-execution-model}) 
considers the influence between nodes and related edges from a control-flow graph to 
assign weights for code blocks \citep{zhang2009a}.
\citeauthor{zhao2010} propose a technique for calculating the suspiciousness of edges 
in a control-flow graph \citep{zhao2010}. This value is used to obtain a weighted coverage for each edge 
that considers the distribution of the control-flow for passing and failing test cases. 
The suspiciousness of each block is then calculated using the passing 
and failing weighted coverages for each block through a \textit{Tarantula-like} metric.

\citeauthor{yu2011a} propose a technique that uses an \textit{Ochiai-like} metric to 
calculate the similarity of control and data dynamic dependences \citep{yu2011a}. This similarity 
indicates the correlation between these dependences and an incorrect 
program behavior. 
The final suspiciousness scores are obtained by considering the maximum value 
between the suspiciousness of control and data dependences for each statement.
\citeauthor{zhao2011a} proposed a technique that uses the influence of edges (branches) in an execution to obtain a  
list of suspicious blocks \citep{zhao2011a}. First, they calculate the suspiciousness of edges. Afterwards, they calculate the \textit{fault proneness} 
of each edge using the total nodes that \textit{in} in the successor node and total nodes that \textit{out} 
from the predecessor node. The approach indirectly uses the dependencies between basic blocks to measure the 
importance of predecessor blocks in their successor blocks.

For their work on fault localization for field failures, \citeauthor{jin2013} 
proposed three strategies to reduce the number of entities (branches) presented for fault localization \citep{jin2013}: (1) 
\textit{filtering}, which excludes entities at three levels: those executed only by passing executions, branches executed for both 
passing and failing executions, and branches that have other control dependent branches; (2) \textit{profiling}, which 
computes the suspiciousness of program entities by considering the number of times an entity is executed in each execution 
for all executions; and (3) \textit{grouping}, which groups entities that belong to the same code region and have 
the same suspiciousness values. These strategies can in most cases pinpoint the faulty entities 
in the first picks. 

\citeauthor{li2014} proposed a technique that uses information from program structure to improve fault localization \citep{li2014}. They consider 
that some structures can influence their statements, making them more susceptible to be wrongly classified by SFL techniques. 
For example, a faulty statement in a main class can be underestimated because it is always executed by both failing and passing 
runs. Conversely, a non-faulty statement in a \textit{catch} block can be overestimated because it is only executed by 
failing runs. 

Some studies use program slicing because of their ability to hold faulty statements in their resultant 
subsets. However, slicing-based techniques must propose strategies for reducing the amount of returned code.
\citeauthor{zhang2005} evaluated three variations of dynamic slicing for fault localization \citep{zhang2005}: data slicing, 
full slicing, and relevant slicing. Regarding the amount of information returned, relevant slices are larger than full slices, which are in turn 
larger than data slices. The results show that, although data slices returned a reduced amount of statements, in several cases the 
fault statement was not returned. Full slicing returned a large number of statements, with the faulty statement in most cases among them.
Relevant slicing returned a slightly larger amount of statements than full slicing, but the faulty statement was always present.
\citeauthor{wong2006} proposed a technique that uses execution slices and inter-block data dependency for fault localization \citep{wong2006}. 
The technique first computes a dice from one failing and one passing execution. If the fault is not in the dice, 
inter-block data dependency from the failing test case is used to add more information. If the amount of code after adding 
such data dependencies is excessive, the technique uses the distinct blocks presented in another passing execution, with 
respect to the previous passing execution, to generate another dice.

\citeauthor{alves2011} used dynamic slicing and \textit{change impact analysis} to reduce the number of statements in the ranking lists 
provided by fault localization techniques \citep{alves2011}. Three techniques were proposed to obtain this reduction: \textit{T1}, which ranks only 
statements that appear in the dynamic slice; \textit{T2}, which applies a change impact analysis before the dynamic slicing and 
considers all statements in the dynamic slice; \textit{T3}, which is similar to T2, but with only changed statements ranked.
\citeauthor{tiantian2011} proposed a technique that calculates suspiciousness of predicates using SFL \citep{tiantian2011}. F-score is used to 
assign values that indicate the likelihood that predicates will be faulty. From a predicate, the technique generates control-flow 
and data-flow information on demand for each predicate by constructing a procedure dependence graph (PDG) 
for the procedure that includes the predicate. Next, backward and forward slices from this PDG are obtained.

\citeauthor{ju2013} proposed a technique that combines full slices from failing executions with execution slices from passing 
executions \citep{ju2013}. These slices are merged using intersection to obtain a hybrid slice.
\citeauthor{wen2011} proposed a technique that combines SFL and program slicing \citep{wen2011}; it removes all program elements that 
do not belong to any failing executions slices. To calculate the suspiciousness of the remaining elements, they consider the 
execution frequency of each element in each test case and the contribution of an element in a test case. The contribution is 
the percentage in which an element is executed, considering all executed elements.
\citeauthor{neelofar2017} combines dynamic and static analysis for fault localization \citep{neelofar2017}. They categorize statements in several 
categories (e.g., assignment, control, function call). These categories are used to weight suspiciousness scores obtained 
using SFL.

\subsection{Combination of SFL techniques}
\label{sub-combination}

The many techniques based on ranking metrics led to studies that combine different approaches to improve 
fault localization. \citeauthor{debroy2011b} used Tarantula, Ochiai, and Wong3 to propose a 
consensus technique, using the concept of rank aggregation, specifically \textit{Borda's method}, to combine statements 
classified with different values by each previous technique \citep{debroy2011b}. 
\citeauthor{ju2013} proposed a new ranking metric called \textit{HSS} \citep{ju2013}. They combined two ranking metrics, O$^{p}$ and Russel\&Rao, 
by multiplying them to calculate the suspiciousness of program entities. \citeauthor{xie2013a} showed both O$^{p}$ and Russel\&Rao to be 
better-performing ranking metrics \citep{xie2013a} (see Subsection~\ref{sub-comparison}).

Other studies use AI to create or combine previous techniques. 
\citeauthor{yoo2012} built several ranking metrics using \textit{Genetic Programming} (GP) \citep{yoo2012}. 
The GP operators used to create such ranking metrics were simple arithmetic operations: addition, 
subtraction, multiplication, division, and square root operation. To measure how well 
a ranking metric classifies faults, they used the EXAM score evaluation metric 
(see Subsection~\ref{sub-evaluation}) as the fitness function.
Six out of 30 ranking metrics created using GP outperformed other ranking metrics used for comparison.
\citeauthor{cai2012} proposed a technique that uses suspiciousness lists from previous ranking metrics \citep{cai2012}. The lists of 28 ranking metrics were 
used as input. The technique uses a \textit{k-means} algorithm to cluster the statements, using the ranking position obtained by each 
ranking metric as features of the statements.

The technique proposed by \citeauthor{le2015b} combines SFL (Tarantula) and Information Retrieval-based fault localization (IRFL) \citep{le2015b}.
IRFL uses bug reports to generate suspiciousness lists \cite{saha2013}.
Their technique combines program elements from failing test cases with related suspicious words from the bug reports. 
The results are present at method level. 

\subsection{Comparison of metric-based techniques}
\label{sub-comparison}

While some studies use several metric-based techniques to evaluate their proposed 
techniques, others compare techniques to identify which are more effective.
\citeauthor{naish2011} use 33 ranking metrics, including their metrics O$^p$ and O, 
which were proposed for single-fault programs \citep{naish2011}. The ranking metrics O, O$^p$, 
Wong3, \textit{Zoltar}, and Ochiai were more efficient in most cases.  
They also show that several ranking metrics are equivalent, producing the same 
ranking lists.
\citeauthor{debroy2011a} show the equivalence between different ranking metrics by 
comparing and simplifying their formulas \citep{debroy2011a}. These ranking metrics 
(e.g., Jaccard, Ochiai, Sorensen-Dice) classify program entities in the same relative 
position in their ranking lists. Thus, one can avoid using equivalent ranking metrics in future work.

\citeauthor{xie2013a} performed a theoretical comparison between 30 ranking metrics used in fault localization \citep{xie2013a}. 
To make this comparison, they grouped statements classified by the ranking metrics according to statements with respective higher, 
equal, and lower scores than the faulty statement. They identified six groups of equivalent ranking metrics, and seven ranking metrics 
that lack equivalent metrics. They also showed that five ranking metrics (O, O$ˆ{p}$, Wong1, Russel \& Rao, and Binary) perform 
hierarchically better than others for two single-fault sample programs.
\citeauthor{le2013a} evaluated the study of \citet{xie2013a} using well-studied programs, showing that the theoretical comparison does 
not hold for such programs \citep{le2013a}. \citeauthor{ju2017} extended the work of \citet{xie2013a} to evaluate multiple faults \citep{ju2017}.

\citeauthor{ma2014} compared seven ranking metrics for SFL \citep{ma2014}. To model the comparison, they proposed a  \textit{Vector Table Model} that 
represents the possible state of each statement for any program. They assumed that a program has a single 
fault. The results show that O, O$^{p}$,  \textit{D$^{E}(J)$}, \textit{D$^{E}(C)$} are equivalent. These metrics 
outperform \textit{Wong1} and also outperform Jaccard and \textit{Kulczynski1} (which are equivalent).
\textit{D$^{E}(J)$} and \textit{D$^{E}(C)$} were proposed by \citet{naish2013}.
\citeauthor{kim2014} carried out another ranking metric comparison, classifying 32 ranking metrics using clustering \citep{kim2014}. The similarity 
was measured using the normalized suspiciousness values of such ranking metrics to compare their effectiveness, and the metrics 
were clustered in three groups of equivalent ranking metrics. They pointed out that these groups have complementary characteristics, 
each of which has its weaknesses and strengths.

Ranking metrics have also presented divergent results regarding programs with real and seeded faults. The study of \citet{pearson2017} 
shows that the Metallaxis \cite{papadakis2015} and O$^{p}$ ranking metrics perform better for seeded faults, while DStar and Ochiai 
performed better for real faults. Thus, several factors should be considered when using ranking metrics, since their performance 
may vary on different study settings.

\section{Program spectra}
\label{sec-spectra}

Program spectra can represent different levels of program structures, from 
fine-grained to coarser excerpts. The different coverage types include statements, 
blocks, predicates, function or method calls, and spectra proposed in the studies. 
Also, some studies combine different coverage levels. 

\subsection{Types of spectra}
\label{sub-spectra-types}

The program entities most commonly used by SFL techniques are statements. The examination of statements 
may lead a programmer directly to the faulty site. However, a large number of statements 
may need to be examined before the faulty statement is found. 
Several works have used statements as their program entity investigation level 
\cite{jones2007,xie2010a,le2013a}. Other works used block coverage 
\cite{wong2006,zhang2009a,xue2013b}.

Predicates are statements such as branches, return values, and scalar-pairs\footnote{A 
scalar-pair is a relationship between a variable assignment and another same-typed 
variable or constant} \cite{liblit2005}. 
\citeauthor{guo2006} use predicate coverage to compare one failing execution to 
all passing executions, and then to identify the most similar passing execution \citep{guo2006}.
This similarity is measured by comparing the results of predicates (true or 
false) and their execution order. 
The technique generates a report composed of predicates that were executed 
only by the failing execution.
\citeauthor{naish2010} propose a technique that generates predicate information 
from statement coverage to perform fault localization \citep{naish2010}. The authors show that 
predicate coverage provides more information about the execution, 
such as execution results and control-flow data, than statement coverage. 
\citeauthor{zhang2009b} perform a statistical study of the distribution behavior 
of predicates that are relevant to the occurrence of failures \citep{zhang2009b}. 
In their experiments, about 40\% of predicates lack normal distribution.
\citeauthor{chilimbi2009} proposed a technique that uses path profiles (intra-procedural 
code segments) and conditional probability to identify paths that are more likely to be faulty \citep{chilimbi2009}.
Their results showed that paths are more precise than predicates for pinpointing faults.

\citeauthor{dallmeier2005} used method call sequences in their technique \citep{dallmeier2005}. The method call sequences that 
income and outcome in an object are summarized to represent a sequence set per class.
Classes whose sequence sets differ from one failing run to several passing runs are more suspicious of being faulty.
In this work, the authors observed that incoming calls are more likely to contain faults than outgoing calls.
\citeauthor{laghari2015} proposed an SFL technique that also produces suspiciousness lists at class level \citep{laghari2015}.
\citeauthor{mariani2011} present a technique that gathers information about interactions 
(sequences of method calls) between software 
components\footnote{Components in this study of \citet{mariani2011} are 
independent programs used by other programs.} \citep{mariani2011}. 
The technique generates an interaction model of passing test cases. 
This model is compared to interactions from failing test cases to indicate 
suspicious method calls. Other works also used method call coverages 
\cite{yilmaz2008,souza2017}.

New coverage types have been proposed for fault localization. 
\citeauthor{santelices2009} compared the performance of different coverages---statement, branch, 
and dua \citep{santelices2009}. They showed that different faults are best located by different coverage types. 
Three approaches were proposed combining such coverages. One of them (\textit{avg-SBD}),
which calculates the average suspiciousness values of statement, branch, and dua coverages, 
achieved better results. In this same study, the authors used a technique that infers an 
approximation of dua coverage (\textit{dua approx}) using branch coverage data, which 
demands lower execution costs.
\citeauthor{yilmaz2008} proposed a technique based on the concept of time spectra \citep{yilmaz2008}. The technique 
collects the execution times of methods in passing runs to build a behavior model for each 
method. Methods executed in failing runs that deviate from the model are considered more suspected. 
As time is an aggregate value, it can also represent sequences of events.

\citeauthor{masri2010} uses a coverage named \textit{Dynamic Information Flow 
Analysis} (DIFA) for fault localization \citep{masri2010}. DIFA is composed of interactions 
performed in an execution, including data and control dependences from 
statements and variables. These interactions are known as \textit{information flow 
profiles}. 
\citeauthor{xu2013} present the coverage \textit{Key Block Chains} (KBC) \citep{xu2013}. 
Each KBC is composed of only one predicate whose execution result is true, 
known as an atomic predicate. Thus, a KBC may have several sizes, according to the 
code blocks that are executed until a predicate is evaluated as true.
\citeauthor{papadakis2015} use mutants as a new type of coverage for fault localization \citep{papadakis2015}. 
They calculate the suspiciousness of mutated versions using Ochiai. Next, they 
map the mutants for the statements using the location in which they were inserted.

Strategies to reduce coverage data gathered can also contribute to enhancing 
fault localization.
\citeauthor{perez2012} propose a technique that reduces the amount of code coverage 
information by choosing the granularity of the inspected elements that are 
collected \citep{perez2012}. Only the most suspect elements in a coarser level are collected at a 
more fine-grained level. The proposed strategy reduces the average execution 
time for performing fault localization.
The technique presented by \citeauthor{wang2015} performs multiple predicate switching to find 
predicates that are critical for revealing faults \citep{wang2015}. To avoid exponential growth due to the number of predicates, 
they first apply Ochiai to reduce the number of predicates on the switching phase.

\section{Faults}
\label{sec-faults}

In practice, a developer does not know how many 
bugs a program has; it may contain single or multiple faults. 
Simultaneous faults may change the execution behavior, 
causing interferences in the test results and affecting the performance of SFL 
techniques. There are different types of faults, which can present varied characteristics, 
also impacting in SFL techniques' suspiciousness lists. 
We present these concerns regarding faults below.

\subsection{Multiple faults}
\label{sub-single_multiple}

Most Spectrum-based Fault Localization techniques are assessed using programs 
with single faults. However, a few techniques were proposed specifically 
for programs with single faults \cite{abreu2011,naish2011,zhang2014}. 
Most techniques that were assessed using single-fault programs do not specify 
such a limitation. For experimental purposes, in which the faults are already known, 
it is acceptable to suppose that a program has only one fault. However, 
it is impossible to foresee how many faults a program has in 
industrial settings. Concerning this 																															, SFL techniques started to be 
evaluated in programs with multiple faults.
The following techniques were proposed addressing multiple faults or using 
multiple faults in their evaluation.

To deal with multiple faults, \citeauthor{zheng2006} used a bi-clustering process to classify 
predicates and failing runs \citep{zheng2006}: they calculated the conditional probability of predicates related to failing runs. The 
selected predicates were applied in a collective voting process, in which each failing run votes for its favorite predicate. 
This voting process is iterative, aiming to identify strong relationships between predicates and failing runs.

Another strategy is to identify similar failing test cases.
\citeauthor{jones2007} introduced a technique to parallelize debugging for multiple faults 
by grouping failing test cases that are most similar to a particular fault \citep{jones2007}. 
They use \textit{agglomerative hierarchical clustering} to obtain these groups, 
called \textit{fault focusing clusters}.
\citeauthor{digiuseppe2012a} investigated factors that can affect failure clustering \citep{digiuseppe2012a}. They show 
that execution profiles and test case outputs can be accurately used for failure clustering. They also verified 
that failures whose profiles and outputs differ due to the presence of multiple faults can impair the failure 
clustering process. Such failures may be pre-processed to improve failure clustering.
In the following study\citep{digiuseppe2012b}, \citeauthor{digiuseppe2012b} presented a failure clustering technique that uses semantic information 
from source code to improve the failure clustering process. 

\citeauthor{liu2008} compared six failure proximity techniques \citep{liu2008}. These techniques are used to classify failure reports 
that are related to the same bugs. A failure proximity technique extracts a failure's signature and uses 
a function to calculate the distance from other failures to group them as related to the same fault. 
\citeauthor{hogerle2014} investigated factors that impact debugging in parallel for multiple faults \citep{hogerle2014}. They highlight the 
infeasibility to obtain pure failure clusters. Such clusters should contain entities and failing test cases entirely 
related to each fault. One trade-off is obtaining clusters that share program entities, 
causing what they called \textit{bug race}. Bug race means that faults can be present in more than one cluster. Another 
trade-off is obtaining clusters that share tests instead of program entities. Bug races are avoided in this case, but some 
clusters may have no faulty program entities.

The work proposed by \citeauthor{dean2009} uses linear programming to locate 
multiple faults \citep{dean2009}. The technique returns a set of statements that explain all 
failing test cases. For the experiments, the single-fault versions of the 
\textit{Space} and \textit{Siemens suite} were joined to create a version 
with all faults per program. \citeauthor{naish2009} conducted experiments with 
several ranking metrics in programs with one and two faults per version \citep{naish2009}. The 
results show that some ranking metrics were more effective for programs with 
two faults, while others were more effective for single-fault programs 
(see Subsection~\ref{sub-heuristics}).

\citeauthor{steimann2009} proposed a technique that uses only failing test cases, assuming 
that in each of them there is at least one program entity (method) that explains the fault \citep{steimann2009}. All entities in these test cases 
have the same probability of being faulty. All possible combinations of methods through the test cases are verified, 
and only combinations that can explain the fault are kept. Some entities can be more present than others in the 
remaining explanations. Thus, the entities are classified according to their frequencies in the explanations.

\citeauthor{abreu2011} proposed a technique for multiple faults called Zoltar-M \citep{abreu2011}. 
The authors used MBD (Model-based Diagnosis) along with program spectra information to obtain groups of 
statements that relate to the existing faults. The MBD is constructed 
under logical propositions from the results of the dynamic execution information. 
\citeauthor{yu2015} presented a technique for distinguishing test cases that fail due to single faults 
from those that fail due to multiple faults \citep{yu2015}. The technique creates test sets composed of one failing 
test case and all passing test cases. Then, they calculated the distance between a failing test case 
and its most similar passing test case. With the presence of multiple faults, tests with large 
distances are more likely related.

SFL techniques that cope with multiple faults often add more complexity to the fault localization 
process in order to isolate such faults. Conversely, the study of \citeauthor{perez2017} investigated 
bug fixing commits from 72 open source projects, showing that 82.5\% of the commits have a single 
fault \citep{perez2017}. This fact may alleviate the need for more complex techniques to deal with multiple faults. 
However, unknown faults may exist in such commits.

\subsection{Fault interference}
\label{sub-interference}

The occurrence of multiple faults may impair fault localization results, which led to concerns about 
interferences between simultaneous faults. 
\citeauthor{debroy2009} showed that simultaneous faults can cause interferences 
in which some faults hide the incorrect behavior of other faults \citep{debroy2009}. Conversely, some 
faults can help to manifest failures related to other faults. The experiments 
performed showed an incidence of fault interference in 67\% of the assessed programs. 

\citeauthor{xue2013b} studied the impact of multiple faults on five fault localization techniques \citep{xue2013b}. 
They showed that fault interference can reduce fault localization effectiveness by 20\% 
using Ochiai. They also demonstrated improved fault localization effectiveness in around 30\%  
of the cases. 
Regarding the ranking metrics used, they observed that some of them, like Tarantula and Ochiai, poorly performed as 
the number of faults increased. The ranking metrics Chi-Square and Odds Ratio exhibited no difference in their performance 
as the number of faults increased.

\citeauthor{digiuseppe2014} investigated the influence of multiple faults on the performance of SFL techniques \citep{digiuseppe2014}. The results showed 
that the presence of multiple faults has little impact (a decrease in 2\% of effectiveness) on the performance of SFL 
techniques to find the first fault. The fault localization effectiveness of the other faults (beyond the first fault) is 
impaired as the number of faults grows. They also showed that the suspiciousness scores of faults tend to decrease as 
the number of faults increases. There were cases of improved effectiveness and other cases in which some 
faults became unlocalizable. They also verified that fault interference occurred in 80\% of the assessed programs.

\subsection{Fault classification}
\label{sub-faults_types}

Few studies have addressed the impact of fault types 
on their techniques. \citeauthor{santelices2009} argue that different 
coverages can contribute to locating distinct fault types, but do not 
present any relationship between the proposed coverages and the fault types 
that such coverages locate better \citep{santelices2009}. \citeauthor{guo2006} assessed their 
technique in the presence of three types of faults: branch faults, 
assignment faults, and code omission \citep{guo2006}.

Faults by code omission are generally difficult to locate \cite{zhang2009a,xu2011a,xie2013a}.
\citeauthor{zhang2007} proposed a technique to deal with faults caused by code omission \citep{zhang2007}. They used the concept of 
implicit dependence to identify indirect dependencies between the use of a variable and a previous conditional statement.

There are few works that show how the proposed techniques deal with specific 
faults. \citeauthor{masri2010} described the faults used in their experiments and 
analyzed the influence of these faults' characteristics on his technique \citep{masri2010}. 
\citeauthor{burger2011} also described the types of faults used in experiments 
and discussed their impact on the new technique \citep{burger2011}. 
\citeauthor{zhang2011a} used a fault classification defined by \citet{duraes2006} 
to verify the frequency of these faults in real programs \citep{zhang2011a}. 

\citeauthor{debroy2010a} pointed out another type of fault that relates to single faults spread over more 
than one statement \citep{debroy2010a}, which is also known as multi-statement faults. 
\citeauthor{lucia2012} examined 374 faults from three real systems to understand when faults are localizable \citep{lucia2012}. By localizable, 
they meant faults present in one or several lines of code in a nearby region, which is the general assumption of fault localization 
techniques. They manually inspected all faults, finding that 30\% of such faults occur in a single line, while around 10\% 
of the faults spread to more than 25 lines each. Less than 45\% of these faults appear in a single method, and less than 
75\% take place in a single file. 
\citeauthor{pearson2017} observed that, from the real faults from the Defects4J database \cite{just2014} (see Section~\ref{sec-programs}), 
76\% are composed of multi-line statements, and 30\% are related to code omission \citep{pearson2017}. There are also 3\% of the faults in non-executable 
code (e.g., variable declaration), which are not covered by SFL techniques. 
\citeauthor{keller2017} found that only 88 of 350 bugs from the AspectJ program (see Section~\ref{sec-programs}) may be identifiable using SFL \citep{keller2017}. 
Besides the cases of misclassified bugs, bugs related to concurrency and environment (e.g., hardware constraints) are difficult to reproduce.
Approaches to automatically classify faults, such as proposed by \citet{thung2012}, can be helpful for evaluating SFL techniques 
in the presence of different fault types.

\section{Programs}
\label{sec-programs}

Several programs have been used to assess SFL techniques; some are program suites 
composed of small programs often used as benchmarks, whereas others are medium and large programs.  
Most of them are open source programs included in software testing repositories, such as 
the Software-artifact Infrastructure Repository (SIR) \cite{do2005}. SIR contains C and Java 
programs prepared for experimental use, including seeded and real faults, and scripts to 
automate the execution of controlled experiments.

Next, we present a description of programs used as benchmarks by the 
SFL techniques. We also describe their characteristics, such as size 
and number of faults.

\subsection{Description of the main programs}

Several programs have frequently been used to carry out fault localization experiments. 
Among them are \textit{Siemens suite} \cite{hutchins1994}, \textit{Unix suite}, 
\textit{Space}, \textit{flex}, \textit{gcc}, \textit{grep}, \textit{gzip}, \textit{make}, 
and \textit{NanoXML}. 
Table~\ref{tb-programs-detail} shows a description of benchmarks often used in the studies, 
the average number of lines of code (LOC) per version, number of faults for all versions, 
number of versions, and the average number of test cases per version. 
The  Siemens suite and Unix suite data show the average of their programs. 
The number of LOC, faults, versions, and test cases may vary throughout the studies. 

\begin{center}
\begin{table}[h]
\centering
\caption{Programs most used by the studies\label{tb-programs-detail}}
\scalebox{0.925}{
\begin{tabular*}{500pt}{@{\extracolsep\fill}llcccc@{\extracolsep\fill}}%
    \hline
    \textbf{Program} & \textbf{Description} & \textbf{LOC} & \textbf{Faults} & \textbf{Versions} & \textbf{Test cases} \\
    \hline
    Siemens suite & 7 programs & 483 & 19 & 1 & 3,115 \\
    Unix suite & 10 programs & 261 & 17 & 17 & 401 \\
    Space & Satellite antenna controller & 6,200 & 38 & 1 & 13,585 \\
    flex & Lexical analyzer & 10,459 & 21 & 6 & 567 \\
    grep & Search for patterns in files & 10,068 & 18 & 6 & 470 \\
    gzip & Data compressor & 5,680 & 28 & 6 & 211 \\
    make & Build manager & 35,545 & 19 & 6 & 793 \\
    NanoXML & XML Parser & 7,646 & 32 & 5 & 216 \\
    Ant & Build manager & 80,500 & 18 & 11 & 871 \\
    gcc & C compiler & 95,218 & 5 & 1 & 9,495 \\
    XML-security & Encrypter & 16,500 & 52 & 3 & 94 \\
    JFreeChart & Chart library & 96K & 26 & -- & 2,205 \\
    Closure Compiler & JavaScript compiler & 90K & 133 & -- & 7,927 \\
    Commons Math & Math library & 85K & 106 & -- & 3,602 \\
    Joda-Time & Date and time library & 28K & 27 & -- & 4,130 \\
    Commons Lang & Text library & 22K & 65 & -- & 2,245 \\
    Mockito & Mocking framework & 11,838 & 38 & -- & 1,854 \\
    \hline
\end{tabular*}}
\end{table}
\end{center}

The Siemens suite comprises seven small programs written in C: 
\textit{print\_tokens}, \textit{print\_tokens2}, 
\textit{replace}, \textit{schedule}, \textit{schedule2}, \textit{tcas}, 
and \textit{tot\_info}. Each program contains one fault per version and 
a test suite including thousands of test cases. The large test suites 
were created to achieve a high testing coverage. 
To simulate realistic faults, the faults were seeded for 
experimental purposes by ten people \cite{hutchins1994}. 
A large number of studies used Siemens suite 
in their experiments \cite{guo2006,dean2009,dandan2014}.
The Unix suite is a collection of Unix utilities written in C that are small 
in size and contain several faulty versions \cite{do2005}. These 
programs are \textit{Cal}, 
\textit{Checked}, \textit{Col}, \textit{Comm}, \textit{Crypt}, \textit{Look}, 
\textit{Sort}, \textit{Spline}, \textit{Tr}, and \textit{Uniq}. The 
faults of Unix suite were seeded using mutation-based injection.
Several works have used Unix suite to assess their techniques 
\cite{wong2010,roychowdhury2012a}. 

Other medium and large-sized programs have been used for fault localization. 
Such programs provide more realistic scenarios for experiments, due to their 
sizes and different domain characteristics.
The Space program, used in several studies, was developed by the 
European Space Agency. It contains 38 real faults discovered during the 
development. The test suite was created by \citet{vokolos1998} and the 
\citet{aristotle2007}, consisting of 13,585 test cases to guarantee that each 
branch is exercised by at least 30 test cases \cite{jones2007}. 
\textit{Flex}, \textit{grep}, \textit{gzip}, and \textit{make} are medium-sized 
Unix utilities also frequently used for experiments \cite{liu2008,wong2012b}.
\textit{NanoXML} is an XML parser for Java with different versions used as 
benchmarks. The program has both real and seeded faults. Ant 
and \textit{XML-security} are other Java programs used 
in SFL experiments \cite{mariani2011,souza2017}.

Other studies have added additional programs for use in experimentation. 
\citeauthor{ali2009} describe the process adopted to prepare the 
\textit{Concordance} program for use in fault localization experiments \citep{ali2009}. 
They also argue that the hand-seeded faults of the Siemens suite 
may not be suitable to represent programs. 
The iBugs project \cite{dallmeier2007} is a repository of bugs that contains 390 from 
three Java programs: AspectJ, Rhino, and JodaTime. Most of the faults belong to 
AspectJ (350), and each of the three programs contains more than 1,000 test cases. The bugs were 
identified in the programs' repositories and have been used in fault localization 
research \cite{lucia2012,keller2017}. 258 of all bugs have an associated test case, which 
can be used for fault localization.

Also, \citeauthor{just2014} created a repository for testing research called Defects4J \citep{just2014}, 
which is currently composed of six large Java programs and 395 real faults. The faults 
were identified and extracted from the programs' repositories. Defects4J has been 
used in several recent studies \cite{le2016,laghari2016,pearson2017}.
In Table~\ref{tb-programs-detail}, projects from JFreeChart to Commons Lang belong to 
the Defects4J repository. \textit{LOC} (in KLOC) and \textit{Test cases} columns contain the values 
reported by the authors \cite{just2014}; for Mockito, which also belongs to Defects4J, 
we obtained LOC using the CLOC program for the latest version available in its repository. 
We also used this version to obtain the number of test cases. 
We do not have the numbers of versions of the Defects4J's programs.

\subsection{Size}
\label{sub-size}

A program's size is often used to refer to it as a large (or real) program, 
or as a small one. There is no precise definition limit to determine a program 
as small, medium, or large in size. \citeauthor{zhang2009a} consider 
flex, grep, gzip, and \textit{sed}---programs that contain between 6.5 and 12.6 
KLOC---as real-life medium-sized programs \citep{zhang2009a}. Other authors consider these 
programs as large in size \cite{debroy2010a,abreu2011}. Space, 
with about 10 KLOC, is deemed a large and real program 
\cite{naish2009,yu2011a}. Generally, we can assume that programs with more 
than 10 KLOC are large programs. Those programs containing between 2 and 10 KLOC are 
medium-sized programs, and programs with less than 2 KLOC are considered small programs.

We identified that most authors consider programs `real' if they are applied to professional use, 
whereas `not real' programs (called \textit{toy} programs) 
are those used for experimental purposes or to perform small tasks, including operational 
system utilities \cite{naish2009,zhang2011a}. Real programs 
in this context also consider the existence of real program faults. We observed 
that the programs considered as medium and large in size by the studies in this 
survey can be assumed as real programs and the small programs as toy 
programs. 

Some experiments with large programs used only a few parts of them.
For the Gcc program, \citeauthor{wong2012b} instrumented one sub-directory (gcc/cp) 
for their experiments \citep{wong2012b}. \citeauthor{mariani2011} uses NanoXML (8 KLOC), Eclipse (17 MLOC), and Tomcat (300 KLOC), 
analyzing interactions occurred in subsets of such programs \citep{mariani2011}. 
Thus, even experiments with programs that vary from millions to hundreds of thousands of 
lines of code may be criticized for their representativeness as large and real programs.

\subsection{Number of faults}
\label{sub-number_faults}

As discussed in Subsection~\ref{sub-single_multiple}, an important concern that should 
be investigated by SFL techniques is the presence of multiple faults. 
To achieve such a goal, programs used in the experiments must also have 
multiple faults. Moreover, some of these faults must change the output behavior of 
other faults to address the study of fault interferences.

The existing programs are generally composed of single faults. Researchers usually 
change the subject programs to generate multiple-fault versions 
for their experiments, yet these modifications can add biases to the evaluation and make 
experiments more difficult to reproduce. 
\citeauthor{jones2007} created 100 versions of Space, each of 
them containing from 1 to 8 faults, randomly combining the single-fault 
versions \citep{jones2007}. 
\citeauthor{abreu2011} also generated multiple-fault versions for gzip, Space, 
and sed in their experiments \citep{abreu2011}. Several other works have used these random strategies 
to generate multiple faults \cite{wang2009,digiuseppe2014}.

Other studies identified faults to carry out their multiple-fault experiments.
\citeauthor{steimann2009} claim that the number of available multiple-fault programs 
is quite limited \citep{steimann2009}. In their experiments, they show an example with three simultaneous 
real faults from the program \textit{Apache Commons Codec}.
\citeauthor{wong2012b} identified five existing faults of the Gcc compiler using the 
\textit{Bugzilla database}. They merged these bugs to create a 5-bug version of 
Gcc for their experiments \citep{wong2012b}.
Identifying real occurrences of simultaneous faults is a time-consuming activity, but it 
can improve the evaluation of SFL techniques.

\section{Testing data}
\label{sec-testing}

The quality of testing data is pivotal for the performance of SFL techniques. Thus, refinements on test suites may impact 
fault localization performance. A desirable characteristic of test suites is the capacity to execute 
distinguishable parts of the code, which can improve the ability of SFL techniques to pinpoint faulty code more precisely.
As large test suites can impact the execution costs of SFL techniques, test suites with reduced size are also desirable.
In this section, we present several strategies that propose improvements in testing data for fault localization.
Concerns regarding evaluation of testing data, coincidental correctness, and use of mutation testing for SFL are also discussed.

\subsection{Test suite improvements}
\label{sub-improvements}

Some works have addressed ways to distinguish program entities between test cases to improve fault localization.
\citeauthor{baudry2006} proposed a testing criterion called \textit{Dynamic Basic Block} (DBB) \citep{baudry2006}, which is a group of statements 
that are executed by the same test cases. These statements always have the same suspiciousness, and are thereby 
indistinguishable: the greater the number of DBBs, the lesser the number of indistinguishable statements, and thus 
the better it is for fault localization.
\citeauthor{hao2010} proposed three strategies to reduce test cases according to their capacity to execute different 
statements \citep{hao2010}. 

Other studies have evaluated the impacts of testing on fault localization and proposed new strategies to improve testing data.
\citeauthor{abreu2009c} observed the influence of test suite quality on SFL \citep{abreu2009c}. They varied the number of test cases that exercised faults 
between passing and failing test cases, and measured the fault localization effectiveness. As expected, having more failing test cases exercising 
faulty statements leads to better effectiveness. However, a limited number of failing test cases suffices. In their experiments, a value of six 
failing runs is optimal, and additional failing test cases do not affect effectiveness.
\citeauthor{xuan2014} proposed a technique for improving testing data used by SFL techniques \citep{xuan2014}. Given a failing test case with more than 
one assertion, they created one test case for each of these assertions. This aims to avoid situations in which an error occurring in a test 
case execution prevents the following assertions from being executed. 
After generating the atomic assertion failing test cases, they apply dynamic slicing to create a list 
of suspicious statements.

Test suite reduction strategies aim to reduce the number of test cases keeping the former coverage level. Thus, 
the execution cost to run the tests reduces, without impacting on the test suite's quality. These strategies are 
especially suitable for regression testing. However, reduced test suite size may impact the effectiveness of SFL techniques.
\citeauthor{yu2008} investigated the effects of test suite size reduction on fault localization \citep{yu2008}. They used ten test suite 
reduction strategies in four ranking metrics. The strategies hold the statement coverage and remove the 
test cases from different outputs (all test cases, only failing test cases, only passing test cases). 
\citeauthor{zhang2015} used category partition to prioritize test cases for fault localization \citep{zhang2015}. 
Program spectra information is not needed for the prioritization---their technique chooses 
test cases with inputs farthest from the previously chosen ones, aiming to obtain a high 
coverage diversity to improve fault localization.
\citeauthor{zhang2017a} applied cloned failing test cases to improve fault localization \citep{zhang2017a}. Their idea was to balance 
the amount of failing and passing test cases.

The occurrence of coincidentally correct test cases and their impacts for fault localization have also been studied in the recent years.
Coincidental correctness can impair SFL techniques by executing faulty entities as passing test cases, reducing their suspiciousness scores.
\citeauthor{masri2014} proposed a technique to identify coincidental correct test cases to improve fault localization \citep{masri2014}. 
They showed that coincidentally correct test cases (CC test cases) are common. They also show that coincidental correctness 
affects SFL techniques by classifying faulty entities with lesser suspiciousness scores. 
The proposed technique uses \textit{k-means clustering} to classify test cases as CC or not. 
\citeauthor{bandyopadhyay2012} extended the previous version of the work of \citet{masri2014}, including interactions with the developer 
to exclude false positive CC test cases \citep{bandyopadhyay2012}. They recalculate the list of remaining suspicious statements throughout the interactions.
Other studies that deal with coincidental correctness for SFL have been proposed \cite{xue2014a,yang2015}.

\citeauthor{guo2015} proposed a technique to evaluate the correctness of test oracles \citep{guo2015}. Since humans act 
as oracles, evaluation mistakes can impair testing and debugging. Their approach considers 
that tests with similar execution traces likely produce identical results. Similar test cases that 
diverge are deemed suspicious.

\subsection{Mutation testing}
\label{sub-mutants}

Mutation testing has been used to propose new SFL techniques. 
\citeauthor{nica2010} proposed a technique to reduce bug candidates by using constraint-based debugging \citep{nica2010}. 
First, statements that do not violate the constraints and that explain the failing test cases are deemed bug candidates.
Second, the technique generates mutants for each bug candidate. Mutants that make the failing test cases pass are used 
to suggest possible faulty sites.
\citeauthor{moon2014} proposed a technique that uses mutation to modify faulty and correct statements \citep{moon2014}. The rationale is that, 
if a mutant inserted in a faulty statement reduces the amount of failing test cases, then the faulty statement is more likely to be faulty. 
Conversely, a mutant inserted in a correct statement which generates more failing test cases is less likely to be faulty. 
\citeauthor{hong2017} proposed a similar approach for multilingual programs \citep{hong2017}.
 
Mutation testing is also used to seed faults for experiments, and to suggest fixes for program repair \cite{weimer2006,debroy2014}.
\citeauthor{ali2009} used mutation testing to generate faults and shown that these faults are similar to real faults \citep{ali2009}.

\section{Practical use}
\label{sec-practical_use}

Spectrum-based Fault Localization's goal is to help developers to find and fix faults. 
For practical use, one needs to understand whether the evaluation metrics that assess SFL techniques reflect what happens in 
development settings. Moreover, the techniques should be assessed by user studies to understand their role in the debugging activity.

In this section, we address concerns related to the practical use of SFL techniques. First,
we present the metrics used to evaluate SFL techniques. We also present experiments 
with developers using SFL techniques in practice. 
Finally, we present the strategies proposed to enrich SFL techniques with contextual 
information.

\subsection{Evaluation metrics}
\label{sub-evaluation}

There are measures often used by studies to evaluate the performance 
of SFL techniques. Fault localization effectiveness is an effort measure which indicates
how much code is inspected using an SFL technique. As most of the SFL techniques generate 
ranking lists, studies often use this approach or a variation of it.

The most commonly used metric is \textit{EXAM score} \cite{chung2008,naish2011,wong2012b}.
This metric represents the developer's effort to find a fault using a 
list of suspicious program entities. The EXAM score is measured as the 
relative position in which the faulty entity was ranked. It represents 
the percentage of entities that must be examined to find the fault.
EXAM score was based on the metric \textit{score}, proposed by \citet{renieris2003}, 
which indicates the percentage of code that does not need to be examined until 
finding a fault. Essentially, EXAM score and score provide 
the same information in inversely proportional way. Several works 
also used score \cite{guo2006,ali2009,zhao2011b}.

There are other metrics similar to EXAM score. \citeauthor{zhang2011a} proposed a 
metric called \textit{p-score} to measure the effectiveness of locating 
suspicious predicates \citep{zhang2011a}. \textit{Expense} \cite{yu2008} is a variation of EXAM 
score for programs with multiple faults: that is, the percentage of code 
verified before locating the first fault. 
To measure the total effort to locate faults for programs with multiple 
faults, \citeauthor{jones2007} propose another variation of EXAM score called 
\textit{total developer expense (D)} \citep{jones2007}, which is the sum of the EXAM score for 
all faults in a program. 
Another metric proposed in this work is \textit{critical expense to a 
failure-free program (FF)}, which measures the time to obtain a 
failure-free program. Assuming that developers work in parallel 
to fix the faults, and for each fault found the program is recompiled, 
FF is the sum of the maximum developer expense at each iteration.

Other metrics identified were \textit{precision} and \textit{recall}, used to 
measure the accuracy of fault localization techniques based on artificial intelligence. 
For the fault localization domain, precision generally means the percentage of entities 
classified by a technique as faults that are in fact faults.
Recall is the percentage of faults correctly classified when considering all faults.
\citeauthor{roychowdhury2011b} used a metric called Metric-Quality to evaluate a technique's ability 
to rank the most important statements with higher values, and the least important 
statements with lower values \citep{roychowdhury2011b}.

Ranking lists commonly classify several program entities with the same suspiciousness scores. 
This fact impacts the evaluation of ranking metrics, which can vary widely.
To deal with ties in ranking lists, \citeauthor{wong2010} measured the best and the worst cases 
for the score metric \citep{wong2010}. The best case considers the fault in the first position of 
the tied entities, while the worst case considers that the fault is in the last position.
\citeauthor{xu2011a} presented a study that shows that ties in SFL ranking lists 
are common \citep{xu2011a}. They propose four tie-breaking strategies to deal with ranking list ties. 

\citeauthor{moon2014} proposed an evaluation metric for fault localization based on \textit{information theory}, called \textit{Locality Information Loss} 
(LIL) \citep{moon2014}. LIL is used to calculate the difference between the true locality and the predicted locality of a fault.
This metric can be applied to any technique that generates ranking lists.

Techniques providing lists of suspicious elements often assume a 
\textit{perfect bug understanding} \cite{hsu2008}, which supposes 
that the developer inspecting a list will immediately identify, understand, and 
fix the fault as soon as s/he reaches the faulty program element. However, this 
may not happen in practice, and the amount of examined code may increase.
As pointed out by \citet{parnin2011}, the measurement of relative positions is quite imprecise. 
The \textit{absolute number} of entities to be inspected before finding a bug can be a more accurate measure, 
regardless of the amount of LOC a program has.
\citeauthor{liblit2005} measured the number of predicates their technique returned \citep{liblit2005}.
\citeauthor{hsu2008} evaluated their technique of bug signatures (see Subsection~\ref{sub-contextual}), measuring the 
absolute number of bug signatures that contain faults \citep{hsu2008}.
Other studies, most of them from recent years, have used the absolute number of inspected entities to 
evaluate their techniques \cite{steimann2009,le2015a,sohn2017,souza2017}.

The evaluation metrics presented here are useful for comparing SFL techniques in experiments. However, user studies with 
developers allow us to verify whether these metrics are a good model of what happens in practice.

\subsection{User studies}
\label{sub-experiments}

Despite the importance of understanding how SFL techniques can be used in practice, 
there are few studies that perform experiments with developers.
\citeauthor{parnin2011} carried out experiments with a group of 
developers using Tarantula \citep{parnin2011}. The authors provided a list of suspicious statements for 
the developers using two programs, each of them containing a single fault. 
The results show that the developers take into account their knowledge 
of the code to search for the faults, and do not usually follow the classification 
order indicated by the SFL technique.
Some other results were observed, including that the perfect bug detection did not occur 
in the experiment. The authors also verified that the position in which 
the faulty statement is classified had no significant impact on the ability of developers to 
find the bugs. The developers suggested improvements, 
such as the aggregation of the results by their classes or files, and 
the provision of input values used in test cases to enrich the debugging.

\citeauthor{perez2013} carried out an experiment with 40 developers to assess their technique for visualization of debugging 
information (GZoltar) \citep{perez2013}. The participants were master's students with more than five years of experience in Java. 
Two groups were formed---control and experimental groups---with 20 students each using the same program and the same fault. 
The experimental group used GZoltar, and all participants were able to locate the bug in seven minutes on average. 
The control group used the Eclipse without GZoltar. Only 35\% of its participants located the fault within the set time 
of 30 minutes.
\citeauthor{perscheid2014} conducted a user study with eight developers to evaluate their program state navigation debugging tool \citep{perscheid2014}. 
All developers were undergraduate students with six years of experience. Four faults 
were debugged by each student, two using the default debugging tool, and two using the new tool. A time limit of fifteen minutes 
was assigned to each fault. In most cases, the developers found the faults using their approach. They also spent less time to 
find the same faults by using the new tool.

\citeauthor{kochhar2016} asked 386 software engineering practitioners about their expectations of issues regarding research in fault 
localization \citep{kochhar2016}. Most participants deemed research in fault localization as worthwhile. Regarding the preferred code granularity level, 
method was preferred by most of them, closely followed by statements and basic blocks. Almost all respondents are willing to adopt 
a fault localization tool that is trustyworth, scalable, and efficient (i.e., a tool that classifies a fault among the top-5 entities 
in most cases). \citeauthor{bohme2017} performed a study regarding the whole debugging process in which 12 professional developers 
were asked to manually debug 27 real bugs \citep{bohme2017}. The authors built a benchmark called DBGBench, which includes fault locations, 
patches, and debugging strategies provided by the participants. This benchmark can be used to evaluate fault localization techniques.

Two recent works replicate the study of \citet{parnin2011}. \citeauthor{xie2016} evaluate the use of SFL with 207 undergraduate 
students and 17 faults \citep{xie2016}. The programs used are classic Computer Science algorithms, with at most 500 LOC. Their results show 
that SFL helped to locate faults only when such faults were ranked between the first positions. Moreover, SFL increased the time 
spent to locate bugs in most cases. They also show that the participants started the debugging activities with a brief overview 
of the code before using SFL. \citeauthor{xia2016} conducted a following study with 36 professional developers and 16 real faults 
from open source projects, showing that SFL improves both effectiveness and efficiency \citep{xia2016}. Most developers started the debugging 
tasks using the SFL lists before inspecting the code and the tests. 

These studies have presented divergent results regarding the usefulness of SFL in practice. However, all studies resulted in at 
least some improvements for participants that used SFL, showing that these techniques can be useful in practice. More studies 
will help understand the current SFL techniques and develop new techniques useful in industrial settings.

\subsection{Contextual information}
\label{sub-contextual}

SFL techniques have in general tried to precisely pinpoint the faulty site. Ranking lists often 
contain suspicious elements sorted only by their suspiciousness 
scores. As a result, elements from different code excerpts can be assigned with higher scores, which may lead to
first picks that have no direct relationship among them---for example, statements that do not belong to a same method or class. 
In practice, when a developer searches for a bug, s/he tries to understand the conditions in which the bug occurs. 
Techniques have been proposed aiming to provide more information for fault localization. 
Contextual information in fault localization is associated with strategies that help developers 
understand bug causes \cite{jiang2007}. 

\citeauthor{jiang2007} proposed one of the first techniques for contextualization in fault localization \citep{jiang2007}. 
Their technique selects predicates likely to reveal faults using two machine learning techniques: \textit{Support Vector Machines} (SVM) 
and \textit{Random Forest} (RF). These predicates are correlated using a \textit{k-means clustering} algorithm. 
Predicates with similar behaviors over the executions tend to be related.
The faulty control-flow paths are constructed based on paths exercised by failing executions that traverse these predicates. 
The control-flow paths are composed of correlated predicates that provide a context for understanding faults.

\citeauthor{hsu2008} presented a technique that provides a list of subsequences of elements (branches) \citep{hsu2008}. These subsequences are called 
bug signatures; each of them may contain one or more branches in their execution order. The technique first classifies 
the most suspicious branches. From failing traces, the amount of branches is reduced using a threshold value. They use a 
\textit{longest common subsequence} algorithm to identify subsequences that are present in all failing executions. 
They are then ordered by their suspicious values.
\citeauthor{cheng2009} extended the idea proposed by \citet{hsu2008} using graph mining to present a list of suspicious subgraphs \citep{cheng2009}. 
Graphs of faulty and correct executions are generated to obtain significant subgraphs that differ in the  executions.
The subgraphs can be extracted at two code levels: blocks or methods.

\citeauthor{hao2009} proposed an interactive fault localization technique that follows the manual 
debugging routine \citep{hao2009}. The technique uses the developer's estimation in the fault 
localization process. The technique recommends checkpoints based on the suspiciousness of statements. 
The developer's feedback is used to update the suspiciousness of statements and choose the next checkpoint.

The technique proposed by \citeauthor{robler2012} provides a list of correlated elements likely to be faulty \citep{robler2012}.
It combines SFL with automated test generation, using one failing test case and generating several test cases. 
Only branches and state predicates, called facts, executed in failing test runs are suspected 
as relevant to faults. Conditional probability is used to estimate the relevance of facts to explain a bug.

Information from source code and code structures are also used to provide contextual information.
\citeauthor{digiuseppe2012c} utilize semantic information for fault localization \citep{digiuseppe2012c}: comments, class 
and method names, and keywords from the source code. The program is instrumented and the source code is parsed to extract 
the information. Terms from the source code are normalized and correlated with the program entities they belong to.
A list of top terms is presented as an outcome. 
\citeauthor{souza2017} use integration coverage (i.e., pairs of method calls) for SFL \citep{souza2017}. They provide two entity-levels to 
search for faults. The first level is a list of suspicious methods named roadmap. For each method, it is possible to inspect the most 
suspicious blocks that belong to it. Two filtering strategies are then used to limit the number of blocks to be checked for each method, 
avoiding the inspection of blocks with lesser suspiciousness scores.

\citeauthor{le2015a} proposed a technique that evaluates the output of SFL techniques (ranking lists) to indicate when this output 
should be used by developers \citep{le2015a}. They used SVM to build an oracle that indicates whether SFL lists are reliable for 
inspection. They identified several features of programs to build the oracle, such as number of failing test cases 
and number of program elements with the same suspiciousness score.

\citeauthor{yi2015} proposed a technique that combines semantic and dynamic analysis to suggest fault explanations 
for regression testing \citep{yi2015}. Semantic analysis is applied to identify statements that cause an assertion to fail.  
Dynamic analysis is then used to identify code changes that retain the failing assertions. These code 
changes are reported as explanations.
The technique presented by \citeauthor{elsaka2015} extracts subsequences of statements from a set of failing executions \citep{elsaka2015}.
These subsequences derive from common subsequence graphs and include variable values from the execution.

\citeauthor{sohn2017} use information of source code metrics along with ranking metrics for fault localization \citep{sohn2017}. 
The code metrics used are related to fault proneness. They apply Genetic Programming and SVM to rank 
the most suspicious methods.
\citeauthor{zhang2017b} proposed the use of the PageRank algorithm \cite{page1999} for SFL \citep{zhang2017b}. First, they classify failing tests 
according to their importance to reveal the faulty code. Tests that execute fewer methods are deemed more important. 
Second, they use static call graph information to verify methods that are connected with other more suspicious methods.
PageRank is then used to calculate the most suspicious methods.

\section{Discussion}
\label{sec-discussion}

In this section, we discuss the main features, results, and challenges of fault 
localization techniques presented in this survey. We follow the structure proposed 
in Figure~\ref{fig-fl_topics} to organize the discussion. 

\subsection{Techniques}
\label{sub-disc-techniques}

Several ranking metrics have been proposed for fault localization, which were 
created or adapted from other areas. Each has its specificity.
Ochiai differs from Tarantula by taking into account the absence of a statement 
in failing runs. Jaccard differs from Tarantula by considering 
statements executed in passing test runs. 
Experiments have shown that Ochiai has presented higher effectiveness 
compared to other ranking metrics \cite{abreu2009c,le2013a,xie2013a}. 
However, the effectiveness of the best ranking metrics is slightly better 
(e.g., around 1\% less code to examine) in most cases, indicating that they 
provide similar results \cite{naish2011,le2013a}. 
This means that there is a ``ranking metric barrier'' for such approach. Can we do 
better to distinguish suspiciousness values from the actual faulty elements?  
A possible way is to investigate whether ranking metrics present better results 
for different types of faults. 

Statistics-based techniques have also been explored by SFL techniques, especially techniques 
using conditional probability. An important issue regarding spectra data is its non-normal 
distribution \cite{zhang2011a}, which indicates that non-parametric techniques can be explored 
for fault localization.

Other ways to enhance fault localization explore program behavior; program dependence and 
artificial intelligence techniques have been used for this purpose. 
However, due to the high computational costs that are inherent to AI techniques, their 
results should be significantly better to compensate such costs. 
SFL techniques that use program dependence information often deal with large amounts of code. 
Thus, it is also necessary an extra effort to develop strategies to reduce information. Such 
strategies may impose high execution costs, especially for large programs. 
Moreover, SFL techniques based on both artificial intelligence and program dependence can 
be used to identify relationships in the internal structures of programs, helping to provide 
contextual information about existing faults. Further studies shall be proposed to provide better 
results.

Combining previous SFL techniques to propose new ones requires deep knowledge of the 
strengths and weaknesses of each technique through different program characteristics, 
which can help to understand new ways to improve fault localization.

\subsection{Spectra}
\label{sub-disc-spectra}

The choice of program spectra influences the performance of SFL techniques,
impacting execution costs, data available to be analyzed, and outputs  for inspection.
Techniques that use coarser spectra data (e.g., method coverage) may have reduced 
execution costs, requiring less code instrumentation. 

Although statement spectrum is more used, searching for faults through single statements 
may be difficult due to the absence of context that isolated statements provide for developers. 
To tackle this issue, grouping most suspicious statements from the same code regions can help to 
understand faults. 
Method spectra may help to bring context to comprehend faults since it is the lowest 
code level that contains the logic of a program. 
Methods were also chosen as the preferred code unit for debugging by developers \cite{kochhar2016}.
However, developers will manually inspect these methods, which may increase the amount of 
code to be verified compared to statements. An increasing number of techniques have used 
method spectra \cite{laghari2016,sohn2017,souza2017,zhang2017b}. However, these studies 
have not yet been compared to them. Class spectra can also help in understanding faults, 
although classes generally contain a large amount of code. 

Different spectra have been combined to improve  fault localization 
\cite{masri2010,santelices2009,yu2011a,le2016}. We note that data-flow information can 
improve fault localization techniques. However, the amount of 
collected data and the execution costs to process it are high compared 
to those based on control-flow. Maybe due to this fact, a few studies have explored 
the use of data-flow for SFL.
Strategies to reduce execution costs to collect data-flow spectra can make these approaches 
more feasible for practical use \cite{araujo2013,araujo2014}.

\subsection{Faults}
\label{sub-disc-faults}

In controlled environments, it may be reasonable to carry out experiments with single-fault programs. 
However, SFL techniques must deal with an unknown number of faults to be adopted by practitioners. 
Most of the proposed SFL techniques have been assessed using single-fault programs.
Several works have evaluated their performance for multiple faults as a complementary study. 
These studies carried out complete experiments for programs with single faults, and 
small experiments with multiple-fault programs \cite{zhang2009a,wong2012a}. In some cases, 
experiments for multiple faults use a reduced number of programs when compared to single-fault experiments. 
SFL techniques have been proposed to deal with multiple faults 
\cite{jones2007,dean2009,xue2013a}. However, their performances have not been compared by these studies.

The absence of programs containing multiple faults makes it difficult to conduct 
experiments. The studies that perform experiments with multiple faults generate their 
multiple-fault versions by randomly merging single-fault versions \cite{jones2007,debroy2009,naish2009}. 
This impairs the comparison between techniques, due to the different 
procedures used to create the benchmarks. 

The presence of multiple faults has been shown to hamper fault localization effectiveness 
\cite{naish2009}. On the other hand, \citeauthor{digiuseppe2014} argued that multiple faults had a negligible 
impact on effectiveness since at least one of the faults is well ranked in most cases \citep{digiuseppe2014}. Their results show that SFL 
techniques had an average 2\% decrease in effectiveness. For large programs, though, 2\% of statements 
may represent a sizable amount of code for inspection. The existence of multiple faults can even increase 
fault localization effectiveness. When two or more faults cause failures on different test cases, which is 
expected for well-designed unit test cases, it is possible that at least one of these faults is well ranked, as 
occurred in the study by \citet{souza2017}.

Another important issue raised by recent studies is the interference between simultaneous 
faults \cite{debroy2009,digiuseppe2014}. More studies are necessary to investigate the 
effects of fault interference on fault localization. Existing studies have already shown 
frequent interference among faults. One concern is that experiments with faults randomly spread 
across programs do not assure that the faults indeed interfere with each other, which may 
result in non-realistic multiple-fault programs.

The behavior of SFL techniques in the presence of different fault types is a 
rarely approached issue. Authors have reported their techniques' difficulties 
in dealing with particular fault types. There are techniques that explicitly 
do not identify faults resulting from code omission \cite{masri2010,yu2011a}. 
Other authors analyzed the behavior of their techniques for specific faults \cite{liblit2005,mariani2011,digiuseppe2012c}.  
By assessing the performance of SFL techniques through different fault types, 
it is possible to understand strengths and weaknesses of such techniques and, 
thus, propose techniques that are more effective to cope with specific fault types. 
It is also possible to improve a technique that does not work well for a fault type 
or even combine techniques that are better for distinct fault types.

To deal with multi-statement faults, SFL techniques must be able to locate faults scattered 
across different code regions. The techniques must also provide hints 
(e.g., names of variables and/or classes) to help developers to locate non-executable 
statements or faulty sites that need additional code.

\subsection{Programs}
\label{sub-disc-programs}

The use of small programs facilitates experiments. It also allows different 
studies to use the same subject programs to compare techniques.
However, the prevalence of small programs or the use of the same benchmark 
programs in experiments impairs the assessment of SFL techniques. 
Thus, the use of several programs from different domains is needed for a comprehensive 
evaluation of SFL techniques.

As discussed in the previous subsection, programs do not contain multiple 
simultaneous faults. Thus, the creation of programs with multiple faults 
can contribute to experimentation in more realistic scenarios. However, creating 
benchmarks for controlled experimentation is expensive and difficult \cite{do2005}.
One way to create benchmarks is to identify faults in software repositories of 
open source programs, as done by \citet{dallmeier2007} and \citet{just2014}. 
This approach provides real faults for experimentation and facilitates reproduction 
of experiments.
Another possible solution to reduce the effort in creating benchmarks is to seed 
faults by using mutation testing, which has been shown to be a good 
simulation of real faults \cite{ali2009}. Conversely, \citeauthor{pearson2017} showed 
that SFL techniques present different results for seeded and real faults \citep{pearson2017}. 

Notwithstanding the current existing benchmarks, new programs should be added to 
increase the diversity of domains and faults for experimentation. For example, 
most of the real programs currently used to evaluate SFL techniques are frameworks 
and libraries. End-user programs should also be assessed to better comprehend the 
performance of SFL. These new programs must also be available to other researchers 
to facilitate their use in future studies.

\subsection{Testing data}
\label{sub-disc-testing}

Software testing is the main source of information for debugging. 
There are several ways to measure test quality. Testing requirements are used to guarantee that the code is widely tested, and most of the 
program elements are executed. \textit{Fault detection rate} is a quality measure used to assess the ability of a test suite to reveal failures. 
Fault localization leads to another desired criterion for test suites: \textit{fault localization capability}. This 
characteristic means that test suites should be able to distinguish program elements from their test cases. 

A natural process for obtaining test suites with higher coverage is to increase their size. However, large test suites lead to greater 
computational costs to execute them. Test suite reduction strategies are then used to minimize the number of test cases without losing 
the ability to failure detection. Moreover, they are also expected to hold the distinctiveness of program elements 
throughout the test cases, i.e., the fault localization capability. Thus, test suite reduction techniques have to cope with a trade-off 
between reducing test size and keeping test cases distinguishable.

Future studies should devise new ways to measure the quality of testing data for fault localization. Test cases that cover a reduced 
amount of code excerpts may be useful to highlight the most suspicious ones. Conversely, individual test cases that cover a large 
code excerpts may impair fault localization by adding an excessive amount of program entities for each execution. 

\subsection{Practical use}
\label{sub-disc-practical}

Evaluation metrics used to measure SFL techniques are based on assumptions that in practice 
may not occur. Measuring fault localization performance by the relative position of a faulty entity 
in a ranking list can mislead the effort to find bugs. For example, if a technique returns the faulty 
entity within 1\% of the most suspicious statements of a program with 100 KLOC, it 
may be necessary to inspect 1 KLOC to find the fault. This may be infeasible in 
practice. Developers tend to leave SFL lists if they do not find
the fault among the first picks \cite{parnin2011}. As pointed out by previous studies 
\cite{parnin2011,steimann2013}, the techniques should focus on the absolute position, with the 
faulty statement among the first positions.
Moreover, perfect bug detection (see Subsection~\ref{sub-evaluation}) does not 
hold in practice, and thus the effort to find faults tends to increase.

The ranking lists provided by SFL techniques are based on suspiciousness 
scores and may be composed of entities with no direct relation to each other 
(e.g., statements that belong to distinct classes). 
This fact may impair the identification of faults. Thus, techniques that provide more 
information can help debuggers understand the conditions in which faults occur. 
Strategies have been proposed to tackle this problem, such as grouping related 
entities, exploring different code levels, adding source code information, or 
presenting subsequences of execution traces. Future approaches should explore new ways 
to reduce the amount of non-relevant information and to enrich the information 
provided to developers.

When developing fault localization techniques, we suppose several assumptions about the 
developers' behavior while performing debugging tasks. However, without 
user studies, one cannot know whether these assumptions hold in practice. Thus, these studies are 
essential for identifying how the techniques are used, to assess the developers' fondness 
of SFL techniques, and to provide guidance on their use in industrial settings. 
Unfortunately, debugging user studies are still scarce and the few existing ones have 
presented divergent results. This may occur due to difficulties to obtain participants for user studies, 
especially professional developers, and samples that are statistically representative. 
Notwithstanding, new user studies are pivotal for understanding the feasibility of SFL adoption 
by practitioners. 

Moreover, there is a lack of evaluations of SFL techniques in real development scenarios 
(e.g., case studies), which can bring findings to improve fault localization. Also, the 
current user studies had participants without previous knowledge of the code being debugged. 
Can SFL improve the debugging performance of developers inspecting their own code?
Future studies should consider these concerns aiming to understand better the use of SFL in practice.

\subsection{Concept map in Spectrum-based Fault Localization}
\label{sub-disc-concept}

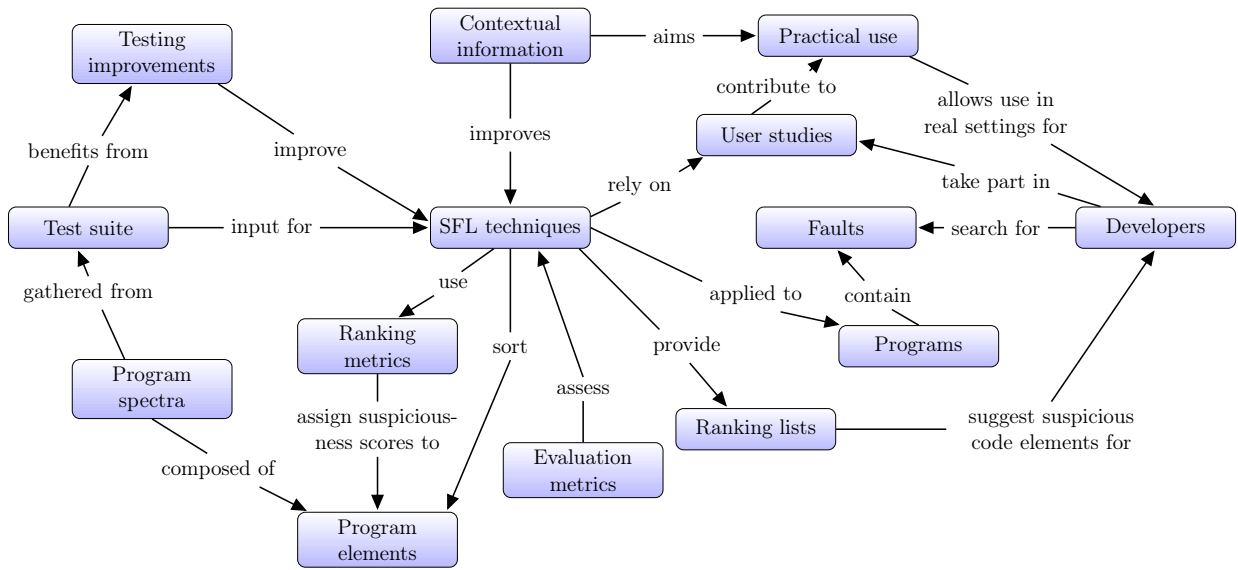
\begin{figure}[h]
\resizebox {\columnwidth}{!}{
\begin{tikzpicture}[node distance=0.25cm,auto]
\tikzset{
  flnode/.style={rectangle,rounded corners,draw=black,text width={width("SFL techniques")+2pt}, text centered,minimum height=0.75cm,top color=white, bottom color=blue!30},
  lknode/.style={draw=none,fill=white},
  lknodelarge/.style={draw=none,fill=white,text width={width("assign suspiciousness")+2pt},text centered},
  edgelabel/.style={midway},
  flarrow/.style={->,>=triangle 45,shorten >=1pt,thick},
  flcaption/.style={text width=7em, text centered},
  flline/.style={-,thick}
}
\node[flnode](sfl){SFL techniques};
\node[lknode,below=0.5cm of sfl](sfl-dummy-south){};
\node[lknode,above=0.25cm of sfl](sfl-dummy-north){};
\node[lknode,below=0.75cm of sfl-dummy-south](sort){sort};
\node[lknode,left=2cm of sfl](test-sfl){input for};
\node[flnode,right=3cm of sfl](faults){Faults};
\node[lknode,right=0.5cm of faults](dev-faults){search for};
\node[flnode,right=0.5cm of dev-faults](dev){Developers};
\node[flnode,left=0.5cm of sort](met){Ranking metrics};
\node[lknode,left=0.5cm of sfl-dummy-south](sfl-met){use};
\node[lknode,left=2.5cm of met](met-dummy-west){};
\node[lknodelarge,below=0.5cm of met](met-elem){assign suspiciousness scores to};
\node[flnode,below=0.1cm of met-dummy-west](spec){Program spectra};
\node[lknode,right=5.75cm of spec](sfl-eval){assess};
\node[lknode,right=2cm of sort](sfl-list){provide};
\node[flnode,right=2cm of sfl-list](bench){Programs};
\node[flnode,below=0.75cm of sfl-eval](eval){Evaluation metrics};
\node[lknode,left=4cm of eval](spec-elem){composed of};
\node[flnode,right=3.5cm of met-elem](list){Ranking lists};
\node[lknode,above=1.75cm of list](sfl-bench){applied to};
\node[flnode,below=1cm of met-elem](elem){Program elements};
\node[flnode,left=1cm of test-sfl](test){Test suite};
\node[lknode,below=0.5cm of test](spec-test){gathered from};
\node[lknode,above=0.75cm of test](test-imp){benefits from};
\node[lknode,right=2cm of test-imp](imp-sfl){improve};
\node[lknode,above=1cm of sfl](cont-sfl){improves};
\node[flnode,above=1cm of cont-sfl](cont){Contextual information};
\node[lknode,right=1cm of cont](cont-pract){aims};
\node[flnode,right=1cm of cont-pract](pract){Practical use};
\node[lknode,right=0.5cm of sfl-bench](bench-faults){contain};
\node[flnode,above=5cm of spec](imp){Testing improvements};
\node[flnode,right=2.5cm of cont-sfl](user){User studies};
\node[lknode,right=1.5cm of sfl-dummy-north](sfl-user){rely on};
\node[lknode,above=0.25cm of user](user-pract){contribute to};
\node[lknode,above=0.25cm of dev-faults](dev-user){take part in};
\node[lknodelarge,above=1.25cm of dev-faults](pract-dev){allows use in real settings for};
\node[lknodelarge,right=2cm of list](list-dev){suggest suspicious code elements for};

\draw[flline]([xshift=-0.3cm]sfl.south) to ([xshift=0.15cm]sfl-met.north);
\draw[flarrow]([xshift=-0.35cm]sfl-met.south) to ([xshift=0.35cm]met.north);
\draw[flline](sfl.south) to (sort.north);
\draw[flarrow](sort.south) to ([xshift=1.25cm]elem.north);
\draw[flline](met.south) to (met-elem.north);
\draw[flarrow](met-elem.south) to (elem.north);
\draw[flline](spec.south) to ([xshift=-0.2cm]spec-elem.north);
\draw[flarrow]([xshift=0.75cm]spec-elem.south) to ([xshift=-1.25cm]elem.north);
\draw[flline]([xshift=-0.5cm]spec.north) to ([xshift=0.25cm]spec-test.south);
\draw[flarrow](spec-test.north) to ([xshift=-0.2cm]test.south);
\draw[flline]([xshift=-0.35cm]test.north) to (test-imp.south);
\draw[flarrow]([xshift=0.25cm]test-imp.north) to ([xshift=-0.35cm]imp.south);
\draw[flline]([xshift=1.2cm]imp.south) to ([xshift=-0.3cm]imp-sfl.north);
\draw[flarrow]([xshift=0.5cm]imp-sfl.south) to ([yshift=0.1cm]sfl.west);
\draw[flline](test.east) to (test-sfl.west);
\draw[flarrow](test-sfl.east) to (sfl.west);
\draw[flline](cont.south) to (cont-sfl.north);
\draw[flarrow](cont-sfl.south) to (sfl.north);
\draw[flline](cont.east) to (cont-pract.west);
\draw[flarrow](cont-pract.east) to (pract.west);
\draw[flline]([yshift=0.2cm]sfl.east) to ([xshift=-0.35cm]sfl-user.south);
\draw[flarrow]([yshift=0.25cm]sfl-user.east) to ([xshift=-1.3cm]user.south);
\draw[flline]([xshift=-0.45cm]user.north) to ([xshift=-0.1cm]user-pract.south);
\draw[flarrow]([xshift=0.6cm]user-pract.north) to ([xshift=-0.25cm]pract.south);
\draw[flline]([xshift=1.3cm]pract.south) to ([xshift=-0.5cm]pract-dev.north);
\draw[flarrow]([xshift=1cm]pract-dev.south) to (dev.north);
\draw[flline]([xshift=-1cm]dev.north) to ([yshift=-0.2cm]dev-user.east);
\draw[flarrow]([xshift=-0.75cm]dev-user.north) to ([yshift=-0.1cm]user.east);
\draw[flline](dev.west) to (dev-faults.east);
\draw[flarrow](dev-faults.west) to (faults.east);
\draw[flline](list.east) to (list-dev.west);
\draw[flarrow](list-dev.north) to (dev.south);
\draw[flline](bench.north) to ([xshift=0.25cm]bench-faults.south);
\draw[flarrow]([xshift=-0.25cm]bench-faults.north) to (faults.south);
\draw[flline](sfl.east) to ([yshift=0.25cm]sfl-bench.west);
\draw[flarrow]([xshift=0.45cm]sfl-bench.south) to ([xshift=-1.4cm]bench.north);
\draw[flline]([xshift=1.25cm]sfl.south) to ([xshift=-0.5cm]sfl-list.north);
\draw[flarrow]([xshift=0.05cm]sfl-list.south) to ([xshift=-0.5cm]list.north);
\draw[flline](eval.north) to (sfl-eval.south);
\draw[flarrow](sfl-eval.north) to ([xshift=0.5cm]sfl.south);

\end{tikzpicture}
}
\caption{Spectrum-based Fault Localization Concept map}
\label{fig-fl_cmap}
\end{figure}

Concept maps \cite{novak2008} are used to organize and represent knowledge. 
Based on the analysis process carried out throughout this survey, we created 
a concept map representing an overview of the main features and their relationships 
within the spectrum-based fault localization area, as shown in Figure~\ref{fig-fl_cmap}. 

Throughout this survey, we present and discuss several advances and difficulties 
of SFL studies. Figure~\ref{fig-fl_cmap} presents the multiple challenges that should be tackled 
to improve SFL for its main objective: to be used in real development settings. 
Through this concept map, researchers can identify their contributions to the area, the issues 
that a study address, how studies relate to each other, and future research topics. 
Several advances were obtained by studies over the period of this survey. We now summarize our 
discussions of this survey from the perspective of the concept map.

\subsubsection*{Summary of advances and challenges}

SFL techniques have proposed several ranking metrics, and it seems that there is not a best 
ranking metric for all scenarios. Some of them achieve better results for different scenarios 
(e.g., single-fault programs, real faults, specific programs). Studies that compared several 
ranking metrics have shown that some ranking metrics produce equivalent results. 
Moreover, the best ranking metrics have achieved small improvements in ranking the faulty 
program elements. Have we reached a limit for the improvement of such metrics? Future work may 
answer this question.

Although the prevalence of statements and branches as the most used program spectra for SFL, several recent 
studies have chosen methods as their program entities. However, only by changing the granularity level is 
not enough to understand whether a spectrum is better. Although method-level spectrum is less precise, it 
may be more comprehensive from the developers' perspective, which can help to understand a bug context. 
Future studies should compare the use of different spectra for fault localization. Moreover, other 
spectra (e.g., data-flow) should be applied to better investigate their usefulness for fault localization. 
User studies are fundamental to evaluate such issues.

The SFL ranking lists are also prevalent in studies. However, other proposals have been presented to provide 
more context for debugging. The examples are small code execution paths and lists combining different spectra. 
How to provide significant information for fault localization and how developers use such information are open 
challenges. 

SFL techniques deal with different types of faults, which can be more easy or difficult to locate. How the 
techniques behave through varied fault types is another open challenge. Faults by code omission have been shown by 
previous studies as a great challenge for SFL techniques. Moreover, interactions among faults have been 
investigated by a few studies and need more attention for a better comprehension of their impact on the 
performance of SFL. 
The characteristics of programs used to assess SFL techniques also impact on their performance and, thus, 
deserve future investigation. In practice, there are programs that differ from most of the benchmark programs 
used in SFL evaluations (e.g., small test suites, large methods, legacy code, multi-language programs). 

Regarding evaluation metrics, recent studies have focused on the absolute number of program 
elements to inspect. Indeed, percentages of inspected code are only useful to compare SFL techniques 
but do not serve to evaluate how techniques will be used in practice since developers do not 
seem to be willing to inspect a large number of code excerpts. 

Testing improvements have also been explored in recent years. These studies have shown that beyond 
code coverage, tests should cover distinguishable code excerpts to improve SFL effectiveness. 

User studies with SFL are rare. However, most of them have been performed recently. 
Each of them has its own experimental design, insights, and limitations. More replications studies 
are needed to better comprehend how developers use SFL. Also, there is a need for studies in real development 
settings, with professional developers, which can bring new findings to understand the practical use of SFL.

Automating debugging is a complex challenge and SFL is a promising approach to locate faults. 
However, SFL is a process that involves: (1) choosing test cases and spectra; 
(2) calculating the suspiciousness of program elements; (3) understanding the strengths and weaknesses 
of an SFL technique through different characteristics of programs and their faults; 
(4) measuring SFL effectiveness; (5) proposing useful SFL outputs; and (6) assessing their practical use. 
By understanding and relating all these concerns, we can propose new ways to improve SFL.

\section{Related work}
\label{sec-related_works}

Other studies were proposed to provide an overview of the fault localization area. 
The studies shown in Section~\ref{sub-comparison} \cite{naish2011,debroy2011a,xie2013a,le2013a} evaluated and compared the 
performance of several ranking metrics. 
\citeauthor{alipour2011} conducted a survey on fault localization \citep{alipour2011}. The author considered that the major fault localization approaches 
are program slicing, spectrum-based, statistical inference, delta debugging, dynamic, and model checking. In his survey, only six 
studies of SFL techniques were addressed---most of the studies are related to model checking-based techniques.
The author concludes that such techniques are far from practical use due to difficulties related to execution time and scalability for 
large programs. Beyond these concerns, model checking techniques usually require formal specifications of programs, which is 
difficult to obtain for most programs.
\citeauthor{agarwal2014} presented a literature review on fault localization, including studies from 2007 to 2013 \citep{agarwal2014}. 
They selected 30 papers from major Software Engineering journals and conferences. Most of the papers are focused 
on test suite improvements for fault localization and SFL techniques. The results are presented in a table describing 
studies' characteristics and a description of the most frequent techniques and strategies in the area.

\citeauthor{wong2016} presented a fault localization survey addressing techniques from 1977 to November 2014 \citep{wong2016}. 
They classified the techniques in eight categories: program slicing, spectrum-based, 
statistics, program state, machine learning, data mining, model-based debugging, and additional techniques. 
Their survey also addresses fault localization tools developed by the presented studies. 

Our survey differs from the previous ones by focusing on spectrum-based techniques from 2005 to October 2017, which includes the most 
recent work of the SFL research area; moreover, we discuss seminal works on automated debugging from the 1950s to 2004 through a historical overview. 
Besides the database search, we applied a snowballing process to extend the searching for fault localization studies.

This survey also differs by addressing topics related to the practical adoption of SFL, such as user studies and techniques that provide 
additional contextual information, aiming to improve developer program comprehension. We understand that 
future research must focus on strategies to allow the use of SFL techniques in real settings.
Moreover, we also included studies that focus on testing improvements and mutation-based techniques for fault localization. 
Another contribution of this survey is to propose a concept map positioning the main topics in the field as well as the relationships among them.

\section{Conclusion}
\label{sec-conclusion}

A great number of fault localization techniques have been proposed in the last decades. 
These techniques aim to pinpoint program entities that are likely to be faulty. Thus, developers 
can inspect these entities to find faults, reducing the time spent debugging.

This survey focuses on spectrum-based fault localization 
techniques, which have presented promising results.
We address the main topics regarding SFL to provide a comprehensive overview of 
the research area: SFL techniques, spectra, faults, programs, testing data, and 
practical use. 

Several advances have been achieved, while some challenges and limitations 
should be tackled to improve SFL techniques.
Among the advances, recent techniques have carried out experiments with real programs. 
Other SFL techniques aimed at providing more information to improve fault localization. 
Moreover, more studies have focused on pinpointing the faults among the first picks 
of the suspiciousness lists.

There are several challenges that future studies should consider. 
New ways for exploring program spectrum information and new strategies for 
generating reduced sets of suspicious entities can contribute to improving the output results.  Combining 
different spectra (e.g., data-flow and control-flow spectra) seems to fine-tune the fault localization ability of SFL
techniques; however, sophisticated spectra  are costly to collect. Strategies for collecting fine-grained coverage levels 
from  suspicious coarser levels can help balance execution costs and output precision. New techniques that cope with 
multiple-fault programs are needed to support  fault localization of real programs, in which the number of faults is 
unknown. Large-size programs from diverse domains, containing different fault types and 
multiple faults, will provide realistic scenarios for assessing SFL techniques. 
More user studies will enable a better understanding of how fault localization techniques are used in practice. 

This survey also presents a concept map of SFL, representing the relationships between the main topics and 
challenges for future research.
By presenting the state-of-the-art of SFL techniques, we hope this survey encourages the development 
of debugging techniques that end up adopted by practitioners.

\section*{Acknowledgments}
This work is supported by FAPESP (S\~ao Paulo Research 
Foundation), under grants 2013/24992-2 and 2014/23030-5.

\bibliographystyle{plainnat}
\bibliography{fl_survey}

\end{document}